\documentclass{aa}  

\usepackage{graphicx}
\usepackage{txfonts}
\usepackage{indentfirst}
\usepackage{amsmath}
\usepackage{hyperref} 
\usepackage{amsmath,bm}
\usepackage{booktabs} 
\usepackage{floatrow}
\usepackage{amsmath,bm}

\usepackage[utf8]{inputenc}
\usepackage[english]{babel}

\usepackage{natbib}

\usepackage[toc,page]{appendix}

\usepackage[nottoc]{tocbibind}

\DeclareFloatFont{tiny}{\tiny}
\begin{document} 
    
\title{Impact of planetesimal eccentricities and material strength \\ on the appearance of eccentric debris disks} 


\author{M. Kim\inst{1}
\and
S. Wolf\inst{1} 
\and         
T. L\"ohne\inst{3}
\and 
F. Kirchschlager\inst{1,2}
\and 
A. V. Krivov\inst{3}}

\institute{Institut f\"ur Theoretische Physik und Astrophysik, Christian-Albrechts-Universit\"at zu Kiel, Leibnizstra\ss e 15, 24118 Kiel, Germany\\
\email{mkim@astrophysik.uni-kiel.de}
\and 
Department of Physics and Astronomy, University College London, Gower Street, London, WC1E 6BT, United Kingdom
\and
Astrophysikalisches Institut und Universit\"ats-Sternwarte, Friedrich-Schiller-Universit\"at Jena, Schillerg\"a\ss chen 2-3, 07745 Jena, Germany}

\date{}
       
\titlerunning  {Impact of collisions on the appearance of eccentric debris disks}
\authorrunning {M. Kim et al.} 


  \abstract
   {Since circumstellar dust in debris disks is short-lived, dust-replenishing requires the presence of a reservoir of planetesimals. These planetesimals in the parent belt of debris disks orbit their host star and continuously supply the disk with fine dust through their mutual collisions.}
   {We aim to understand effects of different collisional parameters on the observational appearance of eccentric debris disks. These parameters are the eccentricity of the planetesimal belt, dynamical excitation, and the material strength.}
   {The collisional evolution of selected debris disk configurations was simulated with the numerical code ACE. Subsequently, selected observable quantities are simulated with our newly developed code DMS. The impact of the eccentricity, dynamical excitation, and the material strength is discussed with respect to the grain size distribution, the spectral energy distribution, and spatially resolved images of debris disk systems.}
   {The most recognizable features in different collisional evolutions are as follows. First, both the increase of dynamical excitation in the eccentric belt of the debris disk system and the decrease of the material strength of dust particles result in a higher production rate of smaller particles. This reduces the surface brightness differences between the periastron and the apastron sides of the disks. For very low material strengths, the "pericenter glow" phenomenon is reduced and eventually even replaced by the opposite effect, the "apocenter glow". In contrast, higher material strengths and lower dynamical excitation of the system result in an enhancement of asymmetries in the surface brightness distribution. Second, it is possible to constrain the level of collisional activity from the appearance of the disk, for example, the wavelength-dependent apocenter-to-pericenter flux ratio. Within the considered parameter space, the impact of the material strength on the appearance of the disk is stronger than that of dynamical excitation of the system. Finally, we find that the impact of the collisional parameters on the net spectral energy distribution is weak.}
   {}


   \maketitle


\section{Introduction}

\noindent Debris disks around main-sequence stars contain fine dust grains, bur the lifetime of these dust grains is much shorter than the stellar age (\citealp {Artymowicz}). Thus, these dust grains are not primordial and should be replenished continuously (\citealp {Backman}; \citealp {Artymowicz}). Consequently, replenishment occurs by collisions, which are at least partially destructive, in other words, planetesimals in debris disks must have sufficiently high relative velocities to create fragments with a broad size distribution, extending down to $\mu\rm m$-sized dust grains. \newline
\indent Spatially resolved images of many debris disks show azimuthal asymmetries of surface brightness (e.g., \citealp {Kalas2005}; \citealp {Lagrange2016}; \citealp {Olofsson}; \citealp {Pan}; \citealp {MacGregor}). In (sub)-millimeter wavelength observations, which trace large grains and so the dust parent bodies, such disks often appear as narrow eccentric belts with some offset between the belt center and the star. The most likely explanation for the eccentricities and offsets is secular perturbations by as yet undiscovered planets (e.g., \citealp {Wyatt1999}; \citealp {Faramaz}). Therefore, interpretation of debris disk observations may allow one to constrain parameters of the alleged planets.\newline
\indent Another goal of the analysis of debris disk observations is to reconstruct the hidden planetesimal populations that produce the observed dust by collisions. Unlike planetesimals, dust is subject to stellar radiation forces that make its distribution different from that of the parent planetesimals. To summarize, in order to constrain parameters of the putative planets and dust parent planetesimals, one has to understand combined effects of the dust-producing collisions in the dynamically perturbed belts and stellar radiation forces exerted on dust.\newline
\indent The link between the dust distributions and observable quantities is also important. What is actually observed is radiation scattered or re-emitted by dust, rather than dust itself. Observations of debris disks at a given wavelength are primarily sensitive to dust particles of a comparable size and to certain dust locations~--- those at which dust emission at this particular wavelength is the strongest. As a result, the observational appearance of debris disks does not directly reflect the underlying dust distributions, and the same disk may reveal dissimilar structure and shape when viewed at different wavelengths. All this further complicates the analysis.\newline
\indent In recent years, many studies have been focused on collisions and radiation pressure effects in debris disks (e.g., \citealp {Dominik}; \citealp {Wyatt2005}; \citealp {Kenyon2005}; \citealp {Thebault2003}; \citealp {Thebault2007}; \citealp {Stark2008, Stark2009}; \citealp {Kuchner}; \citealp {Thebault2012}; \citealp {Kral2013}; \citealp {Nesvold2013}; \citealp {Thebault2014}; \citealp {Kral2015}; \citealp {Nesvold2015a, Nesvold2015b}; \citealp {Esposito}). For these studies, N-body codes have been used that follow the trajectories of individual particles by numerically integrating their equations of motion. In our study, we use the kinetic approach of statistical physics, which is based on the continuity-like equation for the distribution of dust in an appropriate phase space, and model the evolution of the phase space distributions of the material (\citealp{Krivov2006}). \newline
\indent The goal of the current study is to answer following questions: What are the impact of selected collisional parameters on the wavelength-dependent observational appearance of debris disks? Is it possible to constrain selected collisional parameters based on specific observational quantities of debris disks? Based on the work by \citet{Loehne2017}, we have conducted studies on planetesimal collisions in debris disks and investigate the dependence of the appearance of debris disks on essential collisional parameters, such as the eccentricity of the parent belt, dynamical excitation of the system (i.e., the dispersion of the eccentricities of parent belt bodies), and the critical specific energy for fragmentation (i.e., material strength) of dust particles. \newline
\indent This paper is organized as follows: Section 2 describes the underlying physical process of collisions in the debris disks. In Section 3, we briefly introduce the applied tools: the collisional code ACE and the code DMS for the simulation of observations. In Section 4, we quantify and discuss impacts of the collisional evolution on observables, in particular, their spectral energy distribution (SED) and brightness asymmetries in spatially resolved images. We summarize our findings in Section 5.\newline
\section{Collisional physics in debris disks}
\noindent The probability of two spherical objects on Keplerian orbits to collide can be expressed as the product of the overlap of the two orbits, the number density of grains at the desired location in the space of orbital elements, the relative velocity and the collisional cross section (\citealp{Krivov2006}). In addition, the collisional outcome depends on many other factors, such as the impact angle, dust species, porosities of dust grains, and hardnesses of projectile and target (Blum \& Wurm 2008). Furthermore, debris disks with larger typical orbital eccentricities can significantly increase the geometrical probability of collision and the collisional velocities (\citealp{Krivov2006}; \citealp{Queck}; \citealp{Wyatt2009}), resulting in an increased efficiency of the collisional cascade. \newline
\subsection{Influence of eccentric belts on collisions}
\noindent Various previous studies discussed the brightness asymmetry in eccentric debris disks, for example, $\epsilon$ Eridani (\citealp{Greaves1998, Greaves2005}; \citealp{Hatzes} \citealp{Poulton}; \citealp{Pan}), HR4796 (\citealp{Wyatt1999} ; \citealp{Schneider2009}; \citealp{Thalmann}; \citealp{Lagrange2012}), Fomalhaut (\citealp{Stapelfeldt}; \citealp{Kalas2005}; \citealp{Pan}; \citealp{MacGregor}), HD 181327 (\citealp{Schneider2006}), HD 15115 (\citealp{Kalas2007}), HD 202628 (\citealp{Krist}), HD 115600 (\citealp{Currie}), HD 106906 (\citealp{Kalas2015}), and HD 61005 (\citealp{Olofsson}), as a result of two competing effects: higher temperatures at the pericenter and higher densities at the apocenter. If the increased emission at short wavelengths at the pericenter exceeds the increased flux due to the over-density at the apocenter, the so-called pericenter glow phenomenon is observed. This effect was detected for the first time in Fomalhaut, a 133AU ring offset by 15AU implying a forced eccentricity of 0.11 (\citealp{Kalas2005}). \newline
\indent On the other hand, \citet{Pan} calculated the lowest order in the time-averaged linear number density of particles as a function of longitude, $f$, in a mild eccentric disks ($e$ $\ll$ 1) ring along the orbit as
\begin{eqnarray}
l_{\rm single}(f)\propto1 - e\cos f. 
\end{eqnarray}
\indent Using a different approach, \citet{Marsh} found a similar enhancement at the apocenter. In summary, since particles orbit faster at periastron than at apastron, their number density in Keplerian orbits decreases at the pericenter and increases at the apocenter by the same fractional amount. \newline\newline
\noindent The ejection condition can be derived from the ratio between the radiation pressure of the stellar radiation-field to the gravitational force, $\beta$ ($\equiv$ $F_{\rm rp}/F_{\rm grav}$). 
\noindent A general expression for the $\beta$ threshold, for which a dust particle is ejected from the system, can be expressed as follows (\citealp{Murray}):
   \begin{eqnarray}
    \beta_{\rm eject} \ge \frac{1}{2}\left (\frac{1 - e^2}{1 + e \cos\phi}\right),
   \end{eqnarray}
\noindent where $\phi$ corresponds to the longitude of the orbit, at which the particle ejected ($\phi$ = 0 and $\phi$ = $\pi$ correspond to periastron and apastron, respectively). Apparently, the $\beta$-ratio, and thus the ejection condition and blowout size, depend on the eccentricity of parent belts. \newline\newline
\noindent The gravitational impact of large planetesimals can be observed, for instance, as higher dynamical excitation (\citealp{Kenyon2004}). Dynamical excitation, which can be characterized by the width of the typical eccentricity $e$ and inclination $i$, determines the relative velocities at which orbits cross. Even though the exact level of stirring in debris disks remains unknown, previous studies of debris disk have shown that stirring models are consistent with observations if a dynamical excitation of up to $\sim$ 0.1 is assumed (e.g., \citealp{Wyatt2007}; \citealp{Loehne2008}; \citealp{Thebault2008}; \citealp{Kennedy}; \citealp{Mueller}; \citealp{Wyatt2012}; \citealp{Loehne2012}).
\subsection{Influence of material strength on the outcome of collisions}
\noindent The second key parameter which determines the outcome of collisions in debris disks is the catastrophic disruption threshold $Q^{\rm *}_{\rm D}$ that is the impact energy per unit of target mass, which is defined as the normalized energy at which the mass of the largest fragment has half of the original target mass (\citealp{Benz}). While for smaller objects this quantity is determined by the material strength, the gravitational binding energy dominates in the case of larger objects ($\ge$\,1km). Thus, the catastrophic disruption threshold $Q^{\rm *}_{\rm D}$ is described by the sum of two power laws (\citealp{Davis}; \citealp{Holsapple}; \citealp{Paolicchi}; \citealp{Durda1997}; \citealp{Durda1998}; \citealp{Benz}; \citealp{Kenyon2004}; \citealp{Stewart}), where we follow \citealp{Loehne2012} and assume a size dependence described by \citet{Benz}, together with the reformulation by \citet{Stewart}, regarding the reduced and the total masses of the colliders:
   \begin{eqnarray}  
       \begin{gathered}
       Q^{*}_{\rm D} = \left [Q_{\rm s} \left(\frac{s_{\rm ij}}{1\rm m}\right)^{\rm -b_{\rm s}} + Q_{\rm g} \left(\frac{s_{\rm ij}}{1\rm km}\right)^{\rm b_{\rm g}} \right] \left(\frac{v_{\rm imp}}{3 \rm km/s}\right)^{\rm 0.5} \\ + \frac{3G(m_{\rm i}+m_{\rm j})}{5s_{\rm ij}}.
       \end{gathered}
   \end{eqnarray}
\noindent Here, the subscripts s and g stand for the material {\bf s}trength and the {\bf g}ravity regimes, respectively. The quantity $Q_{\rm s}$ is relevant for describing shock disruption in the strength regime, while $Q_{\rm g}$ is the corresponding specific energy in the gravity regime, both scaled by the impact velocity $v_{\rm imp}$. We denote the equivalent radii of the spheres, $s_{\rm i}$ and  $s_{\rm j}$, with $s_{\rm ij}$ $\equiv (s^3_{\rm i} + s^3_{\rm j})^{\rm 1/3}$ as the combined volume of the colliders. The last term of Equation~(3) approximates the specific energy required to overcome self-gravity, which is important only for radii $\geq$ 30\,km in our simulations (\citealp{Loehne2012}). We choose $Q_{\rm s}$ = $Q_{\rm g}$ = 5\,$\times$\,10$^2$~J/kg, $b_{\rm s}$~=~0.37, and $b_{\rm g}$ = 1.38 (\citealp{Benz}). \newline
\indent Depending on the masses and the impact energy of the colliders, following outcomes are expected: If the energy is sufficient to overcome both the material strength and the combined gravitational potential of the colliders, disruption and dispersal occur. Collisions with specific energies below this threshold are depicted as cratering if the target retains at least half of its original mass. In both cases, a cloud of smaller fragments is produced from the remnants (\citealp{Loehne2017}). If both colliders stay intact, they are assumed to separate again, unless the impact velocities are below 10 m/s. Only for very low relative velocities ($\leq$ 1-10 m/s) sticking and hence growth may occur (\citealp{Poppe2000}). However, such low collision velocities are very rare.

\section{Numerical models}

\subsection{Dynamical modeling}
\noindent As discussed in Sect. 1, most of the previous studies focused on purely gravitational models of planet-disk interactions and forces using the N-body simulations. Following the trajectories of individual objects is useful when the dynamics are complex, whereas any collisional event is assumed to simply eliminate both colliders (\citealp{Lecavelier}). Even though this method is hardly applicable in the case of a sufficiently large number of objects to cover a broad range of particle masses, a superparticle-method is applicable in the case of shorter collisional timescales or more violent collisional events (e.g., \citealp{Nesvold2013}; \citealp{Kral2013}).\newline      
\indent The kinetic method of statistical physics is more suitable for particle distributions and long-term collisional evolution. The kinetic method introduces a multidimensional phase-space distribution of dust (e.g., a distribution of grain sizes, coordinates, and velocities, and the supply, loss, and transport of dust grains). We apply the collisional code \textbf{ACE} (\textbf{A}nalysis of \textbf{C}ollisional \textbf{E}volution, \citealp{Loehne2017}) which allows the simulation of the dust production and removal in the planetesimal belt as well as the dynamical evolution of the whole disk. ACE derives the velocity distribution in particle ensembles with the Boltzmann equation and the size distribution using the Smoluchowski equation for describing the evolution of the solids, including disruptive and cratering collisions. The code allows the implementation of a three-dimensional kinetic model with masses, semi-major axes, and eccentricities as phase-space variables. Thus, the resulting equation can be solved for the phase-space distribution as a function of time, from which mass, size, and spatial distributions can be calculated (\citealp{Krivov2006}; \citealp{Krivov2008}; \citealp{Krivov2010}; \citealp{Loehne2017}). The applied version of the ACE code (\citealp{Loehne2017}) allows us to consider the radial and azimuthal variations of the disk density distribution. Collisional outcomes are calculated with size-dependent scaling laws for fragmentation and dispersal in both the strength and the gravity regime as mentioned in Section 2.2. A more detailed description of this simulation tool can be found in previous studies (e.g., \citet{Krivov2005, Krivov2006, Krivov2008, Krivov2010}, and \citet{Loehne2008, Loehne2012, Loehne2017}).
\subsection{Simulated observations}
\noindent The results of the dynamical calculations provide the basis for the simulation of the observational appearance of the considered disk models. For this purpose, we apply our newly developed code \textbf{DMS} (\textbf{D}ebris disks around \textbf{M}ain-sequence \textbf{S}tars) which allows us to calculate the spectral energy distribution (SED), wavelength- and inclination dependent scattered light images, and polarization maps as well as thermal re-emission maps. It is based on the approach outlined by \citet{Ertel2011, Ertel2012}. To spatially resolve the local density and temperature distribution sufficiently well, an adaptive mesh refinement method in Cartesian coordinates is used. The optical properties of the dust grains are computed using the tool \textbf{miex} (\citealp{Wolf}) which is based on Mie theory (\citealp{Mie}).\newline
\indent In the current study, the density distribution is provided by ACE in a tabular form. Thermal re-emission and scattered radiation maps are computed at different observing wavelengths for each dust particle location and dust property. A more detailed discussion of the DMS code can be found in Appendix A. 
\subsection{Model parameters}
\noindent The basic debris disk model applied in our study is described in \citet{Loehne2017}. In the following, the basic characteristics of that model are summarized in brief. To quantify and identify the specific influence of collisions on the appearance of debris disks, we model a fiducial idealized typical debris disk system around a Fomalhaut-like star that is an A3V main-sequence star (see Table 1). The stellar photospheric emission is described by the corresponding PHOENIX/NextGen grid models (\citealp{Hauschildt}). \newline
\noindent The particles are assumed to be composed of generic astrosilicates (\citealp{Draine}) and ice (\citealp{Li}) in equal volume fractions, with a bulk density of 2.35\,g\,cm$^{\rm-3}$. We use the effective medium theory (Bruggeman 1935) to calculate the optical properties. Since we consider eccentric debris disks in our model, the blowout limit varies between two extreme orbital points that is periastron and apastron (see Eq. 2). For the given photospheric spectrum and an eccentricity $e_{\rm b}$ = 0.4, the blowout limit amounts to $\sim$ 3\,$\mu\rm m$ and 6\,$\mu\rm m$ at periastron and apastron, respectively. Asymmetries are most easily identified where narrow belts are resolved, which correspond to narrow volumes in the space of orbital elements (\citealp{Loehne2017}). The relative radial width $\Delta q_{\rm b}$/ $q_{\rm b}$ at apastron $q_{\rm b}$ = $a_{\rm b}$(1 + $e_{\rm b}$) can be estimated from 
   \begin{eqnarray}
    \frac{\Delta q_{\rm b}}{q_{\rm b}} = \sqrt{\left(\frac{\Delta a_{\rm b}}{a_{\rm b}}\right)^2 + \left(\frac{\Delta e_{\rm b}}{1 + e_{\rm b}}\right)^2},
   \end{eqnarray}
\noindent where $a_{\rm b}$ and $e_{\rm b}$ are varied independently. Here, the quantities $a_{\rm b}$ and $e_{\rm b}$ are the semi-major axes of a perturbed belt object and average belt eccentricity, respectively. The quantity $\Delta e_{\rm b}$ denotes the dispersion of the eccentricities of parent belt bodies (i.e., dynamical excitation of the system). For a more detailed description of the model setup, we refer to \citet{Loehne2017}.\newline
\begin{table*}
\def\arraystretch{1.2}																
\caption{Model parameters for simulation of the dynamical evolution.}               	
\label{table:1}                                         								
\centering                                             								
\begin{tabular}{l l}                                   								
\hline\hline                                              							
  Parameter                     														& Value \\
\hline
  Stellar type                           											& A3V \\
  Mass of the star $M_{*}$                 											& 1.92\,$ M_{\rm \odot}$ \\
  Luminosity of the star $L_{*}$             										& 16.6\,$ L_{\rm \odot}$ \\
  Effective temperature $T_{*}$             											& 8590\,K (\citealp{Mamajek}) \\
  Radial extension    																& 5\,au $\leq$ R $\leq$ 600\,au \\
  Number of grid points in radial direction 											& 694 \\
  Azimuthal  extension 																& 0 $\leq$  $\phi$ < 2$\pi$ \\
  Number of grid points in azimuthal direction  										& 6284\\
  Size of dust grains and planetesimals $s$          									& [0.264\,$\mu\rm m$, 48.6\,km]  \\
  Initial size distribution $n(s)$     		 										& $n(s)$ $\propto$ $s^{\rm  - 3.66}$$T_{\rm orbit}$/$T_{\rm 0}$  \\
  Belt eccentricities $e_{\rm b}$			        	  								& $e_{\rm b}$ = 0.0, 0.1, 0.2, and 0.4 \\
  Dynamical excitation $\Delta$$e_{\rm b}$			 								& $\Delta$$e_{\rm b}$ = $\pm$ 0.05, 0.1, and 0.2\\
  Catastrophic disruption threshold 	$Q_{\rm D}^{\rm*}$								& "very high" ($Q_{\rm s}$ = 1.25\,$\times$\,10$^{\rm 4}\,$J/kg),\\
  (material strength $Q_{\rm s}$)													& "high" ($Q_{\rm s}$ = 2.5\,$\times$\,10$^{\rm 3}\,$J/kg),\\
   																					& "reference" ($Q_{\rm s}$ = 5\,$\times$\,10$^{\rm 2}\,$J/kg),\\
  																					& "low" ($Q_{\rm s}$ = 1\,$\times$\,10$^{\rm 2}\,$J/kg),\\
  																					& "very low"  ($Q_{\rm s}$ = 0.2\,$\times$\,10$^{\rm 2}\,$J/kg)\\
  Dust composition and references of 												& Homogeneous mixture of \\
  corresponding optical data															& astrosilicate (\citealp{Draine}) and \\
                                        												& water ice (\citealp{Li}) \\
  Bulk density  of dust $\rho_{\rm bulk}$            								& 2.35\,g\,cm$^{\rm-3}$ \\
  Distance to the debris disk system										& 8\,pc\\

\hline                                   											
\end{tabular}
\end{table*}
\indent We consider four different eccentricities of the parent belt, $e_{\rm b}$ = 0.0 (circular), 0.1, 0.2, and 0.4, respectively. In addition, three different levels of dynamical excitation, $\Delta$$e_{\rm b}$ = 0.05, 0.1, and 0.2 are considered. The dispersion is defined by the range of eccentricity values. It is parameterized by the maximum and minimum orbital eccentricities that planetesimal belts have at the onset of the collisional cascade. Furthermore, we consider five different values of the material strength $Q_{\rm s}$, which is the strength component of the catastrophic disruption threshold $Q_{\rm D}^{\rm*}$, ranging from "very high" to "very low" (see Table 1). These are related to our reference value for $Q_{\rm s}$ by $Q_{\rm s}$ = $\alpha$\,$\times$\,reference value of $Q_{\rm s}$, with $\alpha$ = 25, 5, 1, 1/5, and 1/25, respectively. Thus, this parameter is spanning a range of 5$^{\rm 4}$ $\sim$ 625 in our simulations. We note that $Q_{\rm D}^{\rm*}$ can be as small as 10$^{\rm -2}$ $\sim$ 10$^{\rm -4}$ J/kg for meter-sized bodies without self-gravity (\citealp{Bukhari}; \citealp{Whizin})\newline
\indent For each run, we evolve the system for $10^{\rm7}$yr. Although many debris disks are much older, we choose this value as a steady-state is achieved within this time (\citealp{Thebault2007}). For the simulation of observations, we consider all grains up to a maximum grain size of around 1 mm. The contribution of larger particles to the observable quantities is negligible. Surveys at sub-mm wavelengths have shown that for most debris disks mass is typically ranging from $\sim$  10$^{\rm-9}$ to several 10$^{\rm-7}$${M}_{\odot}$ (e.g., Greaves 2005, and references therein). Thus, we consider a collisionally evolved steady-state disk with 0.7\,$\times$\,10$^{\rm-8}$${M}_{\odot}$.  
\section{Results}
\noindent In the following, we analyze impacts of the eccentricity $e_{\rm b}$, dynamical excitation $\Delta$$e_{\rm b}$, and the catastrophic disruption threshold $Q^{\rm *}_{\rm D}$ on the observational appearance of the considered debris disks. For this purpose, we first discuss impacts of these parameters on the resulting particle size distribution (Section 4.1). Subsequently, we analyze the corresponding SED (Section 4.2) and spatially resolved observations (Section 4.3). We focus our studies on the comparison between the periastron and the apastron side of the disk (Section 4.4).

\subsection{Particle size distribution}
\begin{figure*}
\centering
\includegraphics[width=18cm]{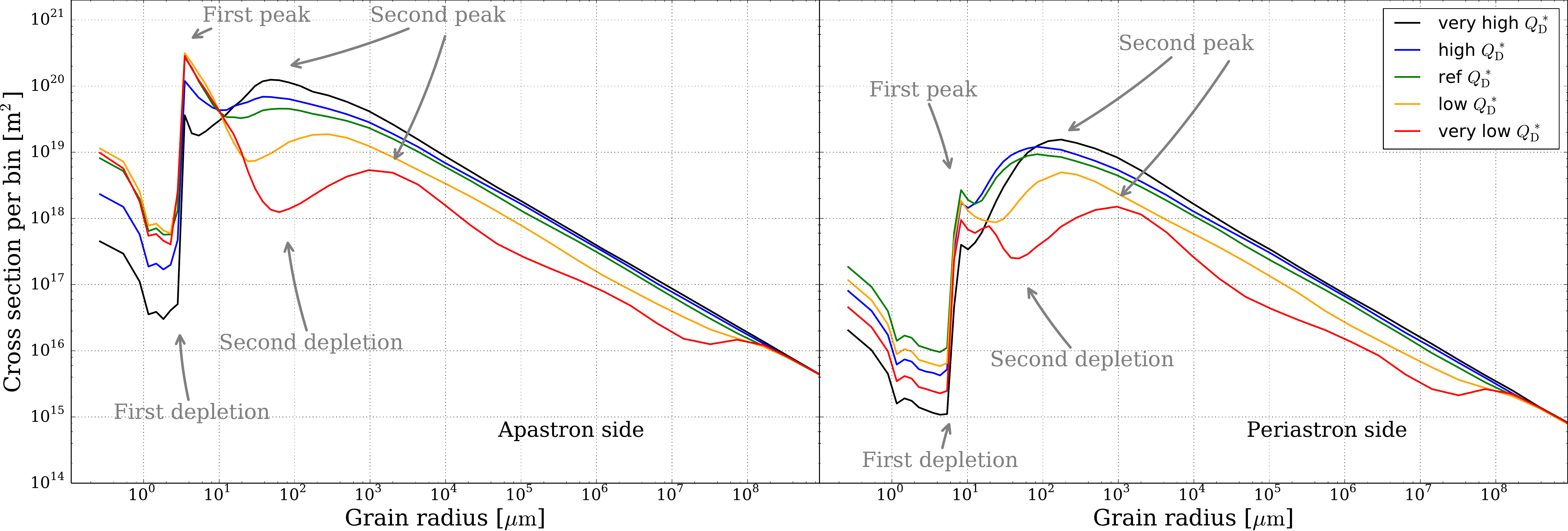}
\caption{The grain size distribution at the apastron (left) and periastron side (right) of debris disks with a belt eccentricity of $e_{\rm b}$ = 0.4 for different material strengths.}
\label{FigVibStab}
\end{figure*}
\begin{figure*}
\centering
\includegraphics[width=18cm]{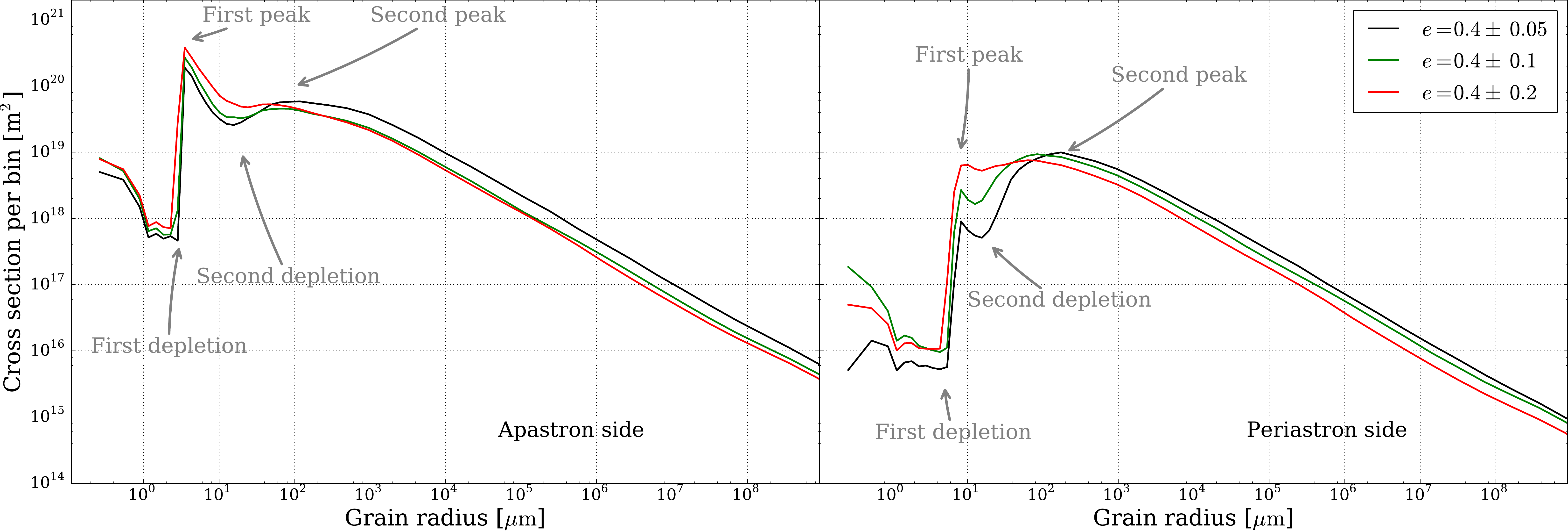}
\caption{The grain size distribution at the apastron (left) and periastron side (right) of debris disks with a belt eccentricity of $e_{\rm b}$ = 0.4 for different levels of dynamical excitation ($\Delta e_{\rm b}$ = 0.05, 0.1, and 0.2).}
\label{FigVibStab}
\end{figure*}
\noindent We first investigate the dust grain size distributions at periastron and apastron of debris disks with different material strengths. As the geometrical cross section of the dust grain is a proxy for the emitted and scattered light, it is shown as a function of the grain size in Figs. 1 and 2. \newline
\indent In Fig. 1, we investigate the dust grain size distributions at the periastron and apastron of debris disks a belt eccentricity of $e_{\rm b}$ = 0.4 (with dynamical excitation $\Delta$$e_{\rm b}$ = 0.1) for different material strengths. Since the radiation pressure has a strong influence on the cutoff in the grain size distribution at smallest sizes, the wavy patterns of the grain size distributions start around the smallest bound grains. We note that the blowout size can be increased with particle porosity and increasing stellar temperature (\citealp{Burns}; \citealp{Kirchschlager}; \citealp{Brunngraeber}). We find different characteristic wavy patterns in the grain size distributions starting with the depletion of grains with radii below the blowout radius $a_{\rm bo}$, which are caused by the lack of $\beta$ > 0.5 dust grains (\citealp{Thebault2003}), depending on different collisional evolutions (Fig. 1 and 2). For grains with radii $\sim$\,3\,$\mu\rm m$ at the apastron side and grains with radii $\sim$\,6\,$\mu\rm m$ at the periastron side, a strong wavy pattern develops. This depletion leads to an over-density of slightly larger grains ("first peak") because grains with radii $a$ < $a_{\rm bo}$ are depleted and thus can no longer contribute efficiently to the destruction and erosion processes anymore. The overabundance of grains with radii around the first peak ($\sim$\,1.5\,$a_{\rm bo}$)  in turn induces another depletion of grains with radii around the "second depletion" ($\sim$\,4 -- 40\,$a_{\rm bo}$) that is caused by small high-$\beta$ grains originating inside the disks. Thus, an efficient destruction is responsible for the deep depletion of objects with radii of~up~to~${\sim\,4\,{-}\,40\,a_{\rm{bo}}}$. Qualitatively, this first wavy pattern is less pronounced in the size distribution of higher material strengths, where the impact velocities and thus their rate of destructive collisions is significantly lower. A strong depletion for lower material strengths is found for grains with radii $\sim$\,60\,$\mu\rm m$, while the depletion in the case of higher material strengths is valid for radii $\leq$\,10\,$\mu\rm m$ (Fig.~1). This depletion eventually leads to an over-density of grains with radii around the "second peak". The overabundance of these grains is shifted from $\sim$\,100\,$\mu\rm m$ (high material strength) to $\sim$\,1000\,$\mu\rm m$ (very low material strength). \newline
\indent These successive domino effects (\citealp{Thebault2007}) propagate the local maximum in the grain size distribution toward bigger sizes and leave a unique characteristic wavy size distribution with a pronounced succession of over-densities and depletions (e.g., \citealp{Campo-Bagatin}; \citealp{Thebault2003}; \citealp{Krivov2006}; \citealp{Thebault2007}). Finally, the amplitude of the depletion of sub-mm grains reaches its maximum for the lowest material strength. As a result, the wavy pattern of the grain size distribution becomes deeper and broader for low material strengths. These results resemble the finding of \citet{Krivov2006} and \citet{Thebault2007}. \newline\newline
\noindent In Fig. 2, we now investigate the dust grain size distributions at the periastron and apastron of debris disks with different levels of dynamical excitation ($\Delta$$e_{\rm b}$ = 0.05, 0.1, and 0.2). Again, we verify that the size distribution converges toward a quasi steady-state wave pattern, whose characteristics depend on the collisional evolution. However, we note that these size distributions (Fig. 2) do not converge toward the same large grain radii ($\sim$\,0.1\,km) for different levels of dynamical excitation $\Delta$$e_{\rm b}$ as for varying the material strength (Fig. 1). If planetesimal belts are more dynamically excited, the number density is -- on average -- decreased. Thus, the ring widths are different and so are the peak densities. \newline
\indent We find that the wavy grain size distribution in highly dynamically excited disks, for instance, in the $\Delta$$e_{\rm b}$ = 0.2 regime, becomes shallower and narrower (see Fig. 2). For the high-excitation case (broad dispersion of the eccentricities of parent belt bodies that is higher $\Delta$$e_{\rm b}$) the depletion of small grains is less pronounced. However, it is propagated to larger grain sizes. Thus, the depletion of sub-mm grains in the case of highly dynamical excited disks is more pronounced which is similar to the case of the low material strength. In contrast, for the low-excitation case (narrow dispersion of the eccentricities of parent belt bodies that is lower $\Delta$$e_{\rm b}$), the depletion of small grains is even more pronounced. Thus, the depletion of sub-mm grains is less pronounced similar to the case of the high material strength. Smaller size particles are strongly affected by the stellar radiation force, which makes their orbits eccentric and their collisional velocities higher. As a result, the collisional lifetimes of small dust grains only weakly vary with the level of dynamical excitation. Thus, the role of destructive collisions for smaller particles becomes less important (\citealp{Thebault2007}; \citealp{Thebault2008}). \newline
\indent We note that this effect results in the second peak in the grain size distribution which is shifted toward larger grains corresponding to a low-excitation case (\citealp{Thebault2008}). Thus the second peak is shifted toward larger grain sizes for smaller values of $\Delta$$e_{\rm b}$ (Fig. 2). In contrast, the rate at which large grains are produced by impacts between much larger planetesimal-sized objects strongly depends on the level of dynamical excitation. Thus, a high dynamical excitation leads to a depletion of grains with radii > 40\,\,$\mu$m and an increased overabundance of smaller grains close to the blowout size (see Fig. 2). Consequently, the final size distribution reflects both effects: a more pronounced depletion of sub-mm grains and a shift of the second peak toward smaller grain sizes with increasing dynamical excitation $\Delta$$e_{\rm b}$. \newline\newline 
\noindent The stirring level in the planetesimal belt and material strength of grains determine the degree of "destructiveness" of collisions (\citealp{Matthews}). Thus, these parameters strongly affect the grain size distribution. Decreasing the material strength has a stronger impact on the grain size distribution than an increased dynamical excitation $\Delta$$e_{\rm b}$ (see Fig. 1 and 2). \newline
\indent In addition, at large grain sizes around 0.1\,km, a change in the slope ("knee") of the size distribution is expected (\citealp{Campo-Bagatin}; \citealp{Durda1998}; \citealp{OBrien}; \citealp{Thebault2007}). This is due to the transition from the strength-dominated regime, where the resistance of colliders to impacts decreases with increasing grain size, to the gravity-dominated regime, where the resistance quickly increases with increasing grain size (Durda \& Dermott 1997). In our simulations, that knee is only pronounced in very destructive circumstances that is for the lowest material strength considered (at $a\,\sim$\,0.4\,km; see Fig. 1). For higher material strengths, the evolution time of 10 Myr was too short to sufficiently converge the size distribution to a steady state. \newline
\indent Since the cratering effect was not included in previous models (e.g., \citealp{Krivov2007}; \citealp{Loehne2008}), wavy patterns were present that were much more pronounced compared to the results from this study. Thus, we can deduce that the cratering impact smoothens wavy patterns. Also, a large fraction of the sub-mm grain depletion is due to cratering impacts (see also \citealp{Thebault2003}; \citealp{Thebault2007}; \citealp{Mueller}). Indeed, small grains of a few $\mu\rm m$ cannot directly break-up objects larger than $\sim$ 1mm (around the second peak), even with their increased impact velocities. However, our results show that they can gradually erode much bigger grains efficiently. \newline
\subsection{Spectral energy distribution (SED)}
\noindent We investigate the influence of different geometries and collisional factors on the resulting SED. We focus on the infrared to the sub-mm wavelength range because the offset from the stellar photospheric SED is largest in this region.\newline\newline
\noindent {\hspace{-2.5mm}} \textbf{1. Dependence on the eccentricity and dynamical excitation} {\hspace{2mm}} \citet{Loehne2017} found that the eccentricity of parent belts affects their SED only weakly. Similarly, we find only a weak influence of the combination of the geometrical offset and collisions on the overall SED for different eccentricities, assuming the reference material strength ($Q_{\rm s}$~=~5\,$\times$\,10$^{\,\rm2}$ J/kg; see Fig.~3). The most noticeable change of the SED is the slight increase of the scattered flux with increasing eccentricity in the near-IR wavelength range. Around a wavelength of 10\,$\mu$m, the difference of fluxes between eccentricities $e_{\rm b}$ = 0.0 and $e_{\rm b}$ = 0.4 reaches its maximum because smaller grains can survive at the apastron side in highly eccentric disks (the different blow-out limits on periastron and apastron side; see Eq. 2). However, the circumstellar dust is hardly detectable through photometric measurements because the stellar photosphere is brighter by a factor of 10$^{5}$ to 10$^{7}$. At wavelengths of $\sim$\,30 - 50 \,$\mu$m, the thermal re-emission of the dust exceeds the direct stellar radiation, increasing by up to 30$\%$ with increasing eccentricity of parent belt. \newline\newline
\noindent Fig. 3 also depicts the influence of different levels of dynamical excitation $\Delta$$e_{\rm b}$. The most noticeable change of the SED is the increase of the scattered flux with increasing the level of dynamical excitation in the near to mid-IR wavelength range. However, again it is hardly detectable through photometric measurements in this wavelength range. Increasing the level of dynamical excitation of eccentric parent belts the disk emission increases by up to 56$\%$ at 30 - 50\,$\mu\rm m$. \newline\newline
\noindent These results are easily explained from the dynamical point of view. Higher eccentricities and an increasing level of dynamical excitation increase the mutual collisional velocities. The efficiency of the collisional cascade is increased resulting in a higher dust production rate. Consequently, the disk appears brighter in those regions where the above effect is strongest. \newline\newline
{\hspace{-1mm}}\noindent \textbf{2. Dependence on the material strength} {\hspace{2mm}} Fig. 4 depicts the influence of the combination of geometrical offset and collisions on the overall SED for different the catastrophic disruption threshold $Q_{\rm D}^{*}$~(material strengths $Q_{\rm s}$~=~1.25\,$\times$\,10$^{\,\rm4}$, 2.5\,$\times$\,10$^{\,\rm3}$, 5\,$\times$\,10$^{\,\rm2}$, 1\,$\times$\,10$^{\,\rm2}$, and 0.2\,$\times$\,10$^{\,\rm2}$~J/kg, fixed eccentricity $e_{\rm b}$ = 0.4). In comparison to the results shown in Fig.~3, variations of the material strength have a more pronounced effect on the SED than the variation of the eccentricity. The flux at optical to mid-IR wavelengths increases with decreasing material strength. Thus the more destructive collisional systems show higher fluxes at wavelengths $\leq$ 80\,$\mu\rm m$. In contrast, the flux at sub-mm wavelengths decreases with decreasing material strength. The deeper and broader wavy pattern of the grain size distribution for low material strengths (see Fig. 1) results in the slope of the SED in this range becomes flatter. However, the SEDs in these wavelength ranges have similar spectral indices which makes it hardly possible to constrain the material strength from the analysis of the SED alone. \newline
\begin{figure}
\centering
\includegraphics[width=9cm]{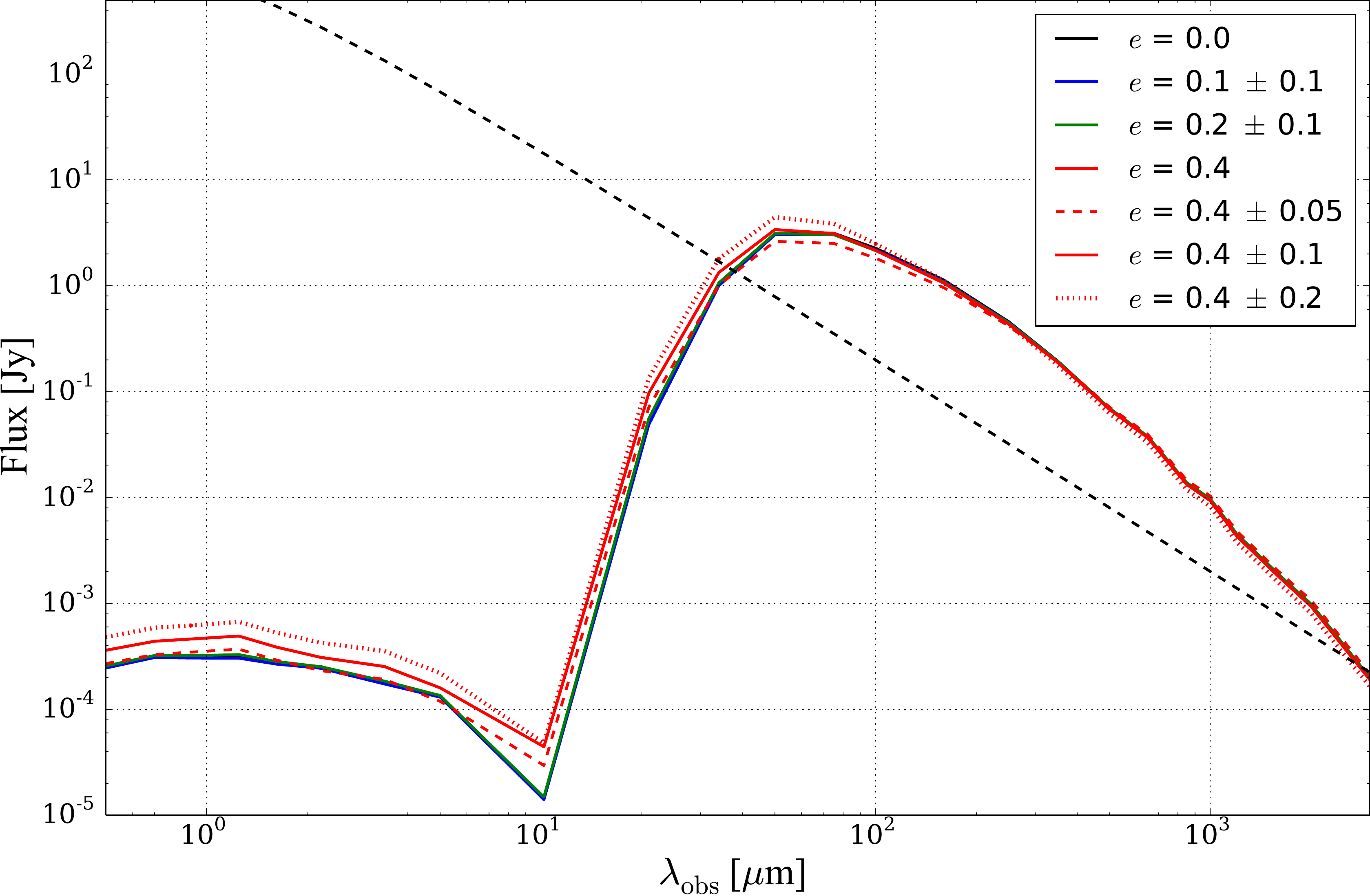} 
\caption{The spectral energy distribution for systems with different eccentricities ($e_{\rm b}$ = 0.0, 0.1, 0.2, and 0.4) and dynamical excitations ($\Delta$$e_{\rm b}$~=~0.05, 0.1, and 0.2). The dashed black line represents the photospheric emission of the central star} \label{FigVibStab}
\end{figure}
\begin{figure}
\centering
\includegraphics[width=9cm]{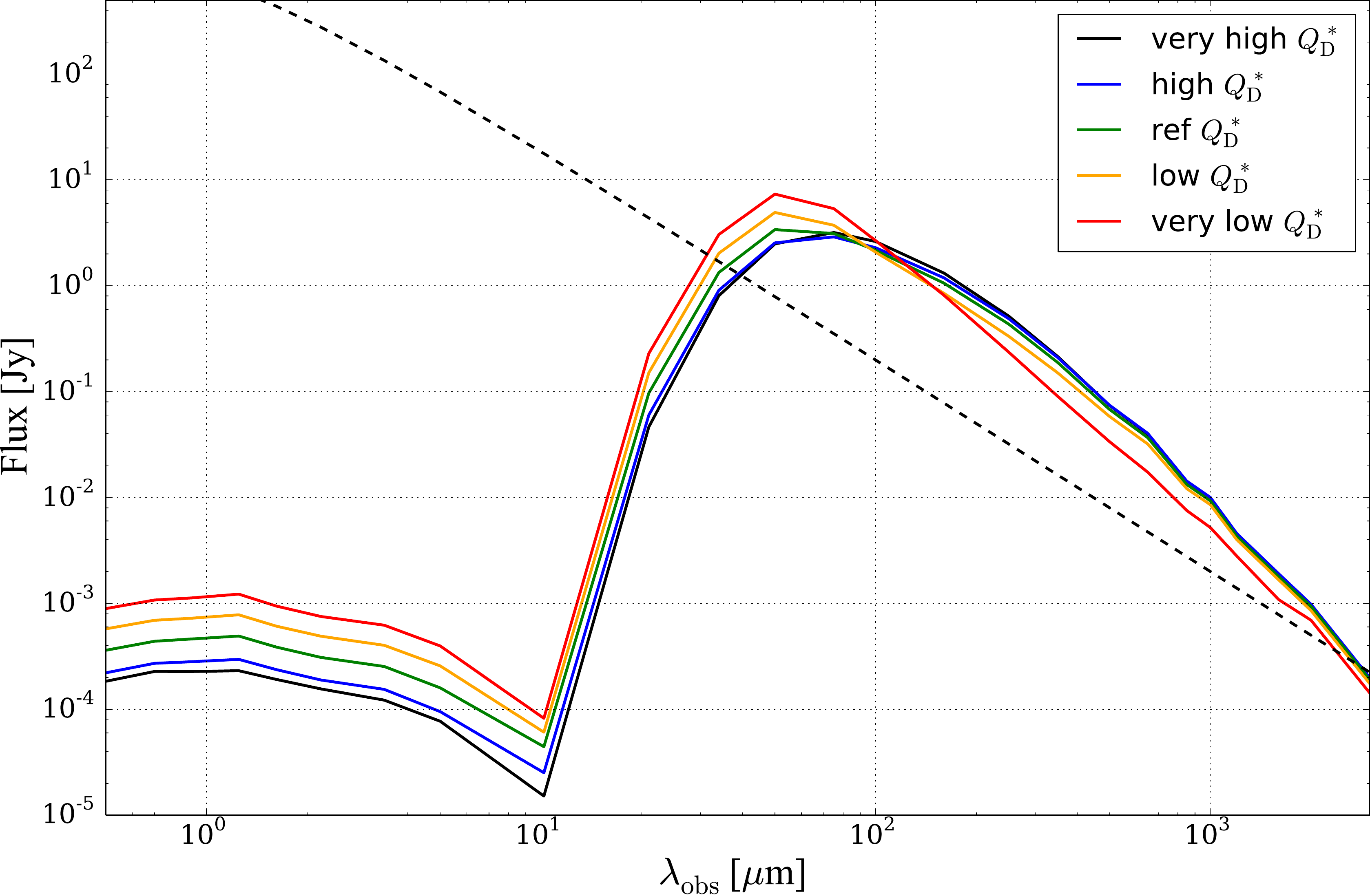}
\caption{The spectral energy distribution for systems with different material strengths. The dashed black line represents the photospheric emission of the central star.}
\label{FigVibStab}
\end{figure}
\begin{table*}
\footnotesize
\def\arraystretch{1.5}
\centering
\caption{A total surface area of the dust in three different size bins resulting from the collisional evolution for different material strengths.}
\label{my-label}
\begin{tabular}{cccccc}
\hline                                
\midrule 
$Q^{\rm *}_{\rm D}$ regime  			& $\dfrac{\rm Surface\,area_{\,\rm total}\,(a_{\rm dust} < 6\mu\rm m)}				    {\rm Surface\,area_{\,\rm total}\,(a_{\rm dust}<1mm)}$ 													& $\dfrac{\rm Surface\,area_{\,\rm total}\,(15\mu\rm m < a_{\rm dust} < 200 \mu \rm m)}	{\rm Surface\,area_{\,\rm total}\,(a_{\rm dust}<1mm)}$ 													& $\dfrac{\rm Surface\,area_{\,\rm total}\,(200\mu\rm m < a_{\rm dust} < 1 mm)}			{\rm Surface\,area_{\,\rm total}\,(a_{\rm dust}<1mm)}$ \\
\midrule 
very high 							& 	39.6 [\%]	&	8.6 [\%]	& 	11.0 [\%]\\
high 								& 	67.8 [\%] 	&	4.1 [\%]	&	6.7 [\%]\\
ref 					       	    & 	72.1 [\%]	&	2.4 [\%]	&	4.9 [\%]\\
low 					      	    &	78.9 [\%]	&	0.5 [\%]& 	2.3 [\%]\\
very low 					     	& 	81.6 [\%]	&	0.2 [\%]	& 	0.7 [\%]\\
\hline                                   
\end{tabular}
\end{table*}
{\hspace{-2mm}}\indent To analyze this finding, we compare the contributions of the dust within different size ranges to the resulting SED, which is relevant for the optical appearance (Table 2). We find a larger surface fraction of grains with radii < 6\,$\mu\rm m$ that is grains which mostly contribute to the SED at optical to mid-IR wavelengths, with decreasing material strength (second column in Table 2). For lower material strengths, the increased production of small grains result in an increase of the SED at these wavelengths. However, we note that surface fraction differences between other collisional regimes are getting smaller because the small grains are more easily affected by stellar radiation. For grains of intermediate size (15\,$\mu\rm{m}\hspace{-1mm}<\hspace{-1mm}a_{\rm{dust}}\hspace{-1mm}<\hspace{-1mm}200\,\mu\rm{m}$; third column in Table 2) that is grains which mostly contribute to the SED at far-IR to sub-mm wavelengths we find the opposite trend. Less destructive collisional systems, corresponding to the higher material strengths, result in higher fluxes in this wavelengths regime. However, because of the marginal change of the slope of the SED in this range, it is not possible to quantify the material strength from pure photometric measurements. The net surface fraction of dust grains with radii > 200\,$\mu\rm m$ is sufficiently large to not to be affected by the stellar radiation pressure (fourth column in Table 2). Thus, collisions become more important here.\newline   
\indent The second most obvious difference of the SEDs of disks with different material strengths is the change of the location of their maximum. They are not only shifted from 50\,$\mu\rm m$ for very low material strengths toward $\sim$ 70\,$\mu\rm m$ for very high material strengths (i.e., for the systems with fewer destructive collisions). They also show an increase of the peak flux with decreasing material strength. \newline\newline
\begin{figure*}
\centering
\includegraphics[width=15cm]{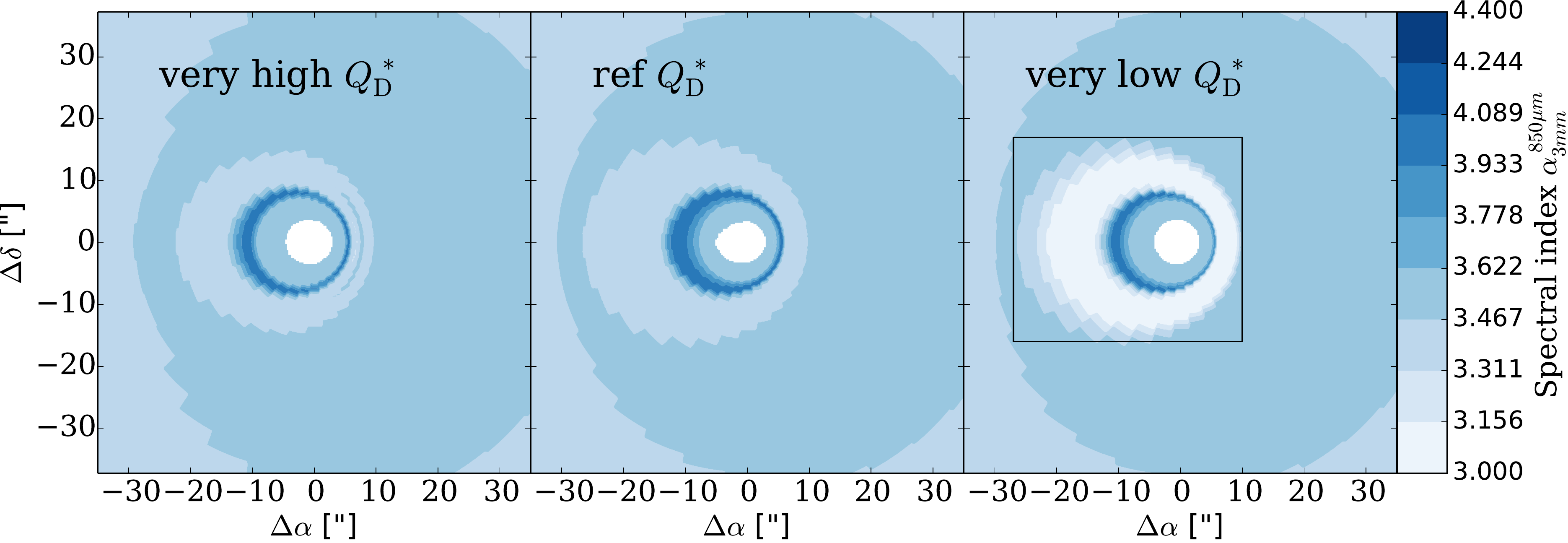}
\caption{The spectral index $\alpha ^{\rm 850\mu \rm m}_{\rm 3mm}$ map for different material strengths ($e_{\rm b}$ = 0.4). We note the low spectral index in the birth ring close to the central star (black square box) in the case of the very low material strength.}
\label{FigVibStab}
\end{figure*}
\subsection{Spatially resolved images}
\noindent In Sect. 4.2, we discussed the dependence between the SED and the varying collisional parameters, for example, dynamical excitation $\Delta$$e_{\rm b}$ and material strength $Q_{\rm s}$. However, as the SED is a representation of net fluxes integrated over the entire disk, we now investigate spatially resolved images and derived quantities, which potentially provide a wealth of additional information. \newline
\subsubsection{Spectral index map}
\begin{figure}
\centering
\includegraphics[width=9cm]{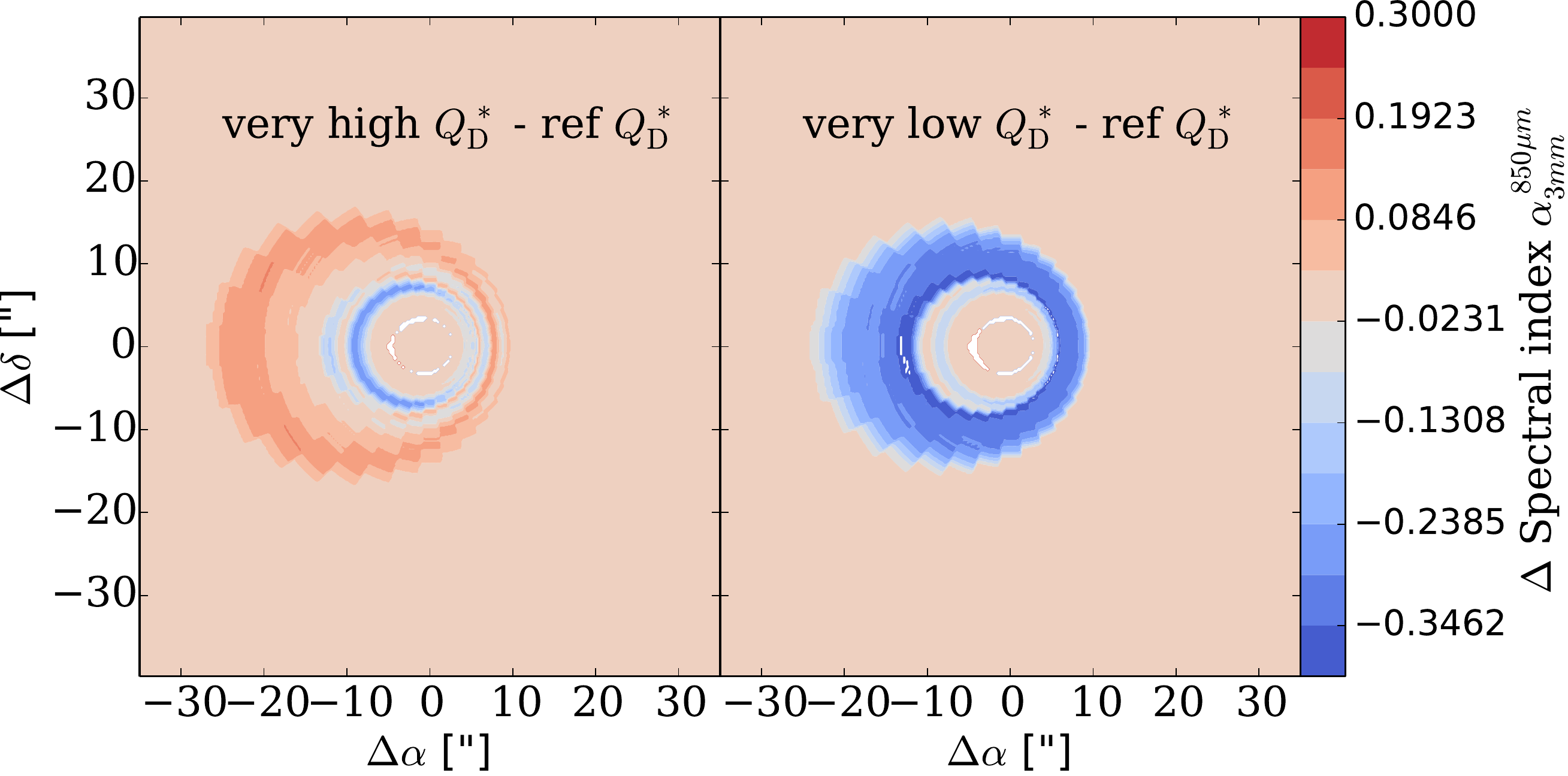}
\caption{Differences between the spectral index $\alpha ^{\rm 850\mu \rm m}_{\rm 3mm}$ maps for the very high material strength (left) and the very low material strength (right) and the reference case (see Figure 5 for comparison).}
\label{FigVibStab}
\end{figure}
\begin{figure}
\centering
\includegraphics[width=9cm]{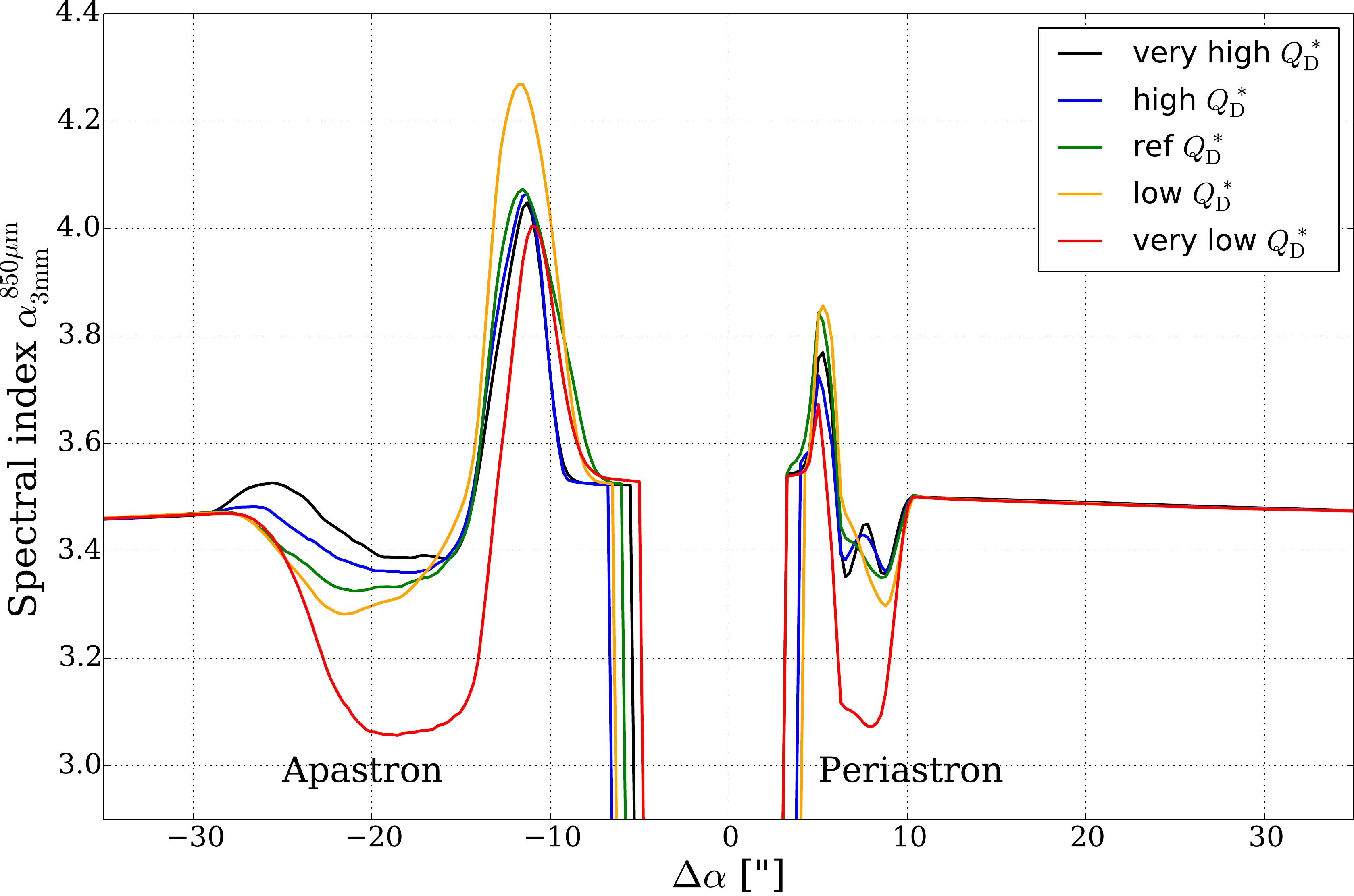}
\caption{The radial cut of the spectral index $\alpha ^{\rm 850\mu \rm m}_{\rm 3mm}$ map for different material strengths ($e_{\rm b}$ = 0.4) along $\Delta\delta$ = 0 (see Figure 5 for comparison). The left and right side of this figure indicates apastron and periastron side of the disks.}
\label{FigVibStab}
\end{figure}
\noindent We begin our analysis with the sub-mm spectral index $\alpha ^{\rm 850\,\mu\rm m}_{\rm\,3mm}$, based on the fluxes at 850\,$\mu\rm m$ and 3\,mm of the observed disks as a function of the material strength for a fixed belt eccentricity $e_{\rm b}$~=~0.4 (see Figs. 5 and 7). In addition, Fig. 6 depicts the difference between the spectral index $\alpha ^{\rm 850\mu \rm m}_{\rm 3mm}$ for the very low and very high material strengths, which indicates the location where large grains (small grains) are more abundant than in the reference case.\newline
\indent We see that the spectral index at larger distances hardly depends on the material strength. Radiation pressure is acting most efficiently on small grains, pushing them to larger distances. Therefore, we find a relatively large fraction of small particles in the outer regions. Consequently, the spectral index is higher at larger distances than around periastron and apastron. Different blowout limits (see Eq. 2), depending on the different locations under which grains are launched from different sites along the parent belts (e.g., around periastron and apastron), shape the local size distributions and eventually result in the asymmetries seen in the halos (\citealp{Lee}; \citealp{Loehne2017}). In consequence, the halo of the periastron side is much more extended than the halo of the apastron side (see Fig. 5 and 7). However, the halo is typically too faint to allow measuring its (sub-)mm spectral index (\citealp{Loehne2017}). \newline
\indent In contrast, the spectral index in the region around the parent belt strongly depends on the material strength. Fig.\,7 illustrates the radial cut of the spectral index around apastron (at right ascension $\Delta\alpha$\,$\sim$\,-20", declination $\Delta\delta$\,$\sim$\,0") and periastron (at $\Delta\alpha$\,$\sim$\,8", $\Delta\delta$\,$\sim$\,0"). One would expect an increased amount of smaller grains in the case of smaller material strengths. However, Fig. 5 (right plot), Fig. 6 (right plot), and Fig. 7 (red line) show the opposite behavior that is a lower spectral index in the case of a very low material strength. At $\lambda_{\rm obs}$ = 850\,$\mu\rm m$, the radiation is dominated by grains with a characteristic grain size (Backman \& Paresce 1993) $a_{\rm c}$\,$\approx$\,$\lambda_{\rm critical}$/2$\pi$\,$\approx$\,145\,$\mu\rm m$ ($\beta$\,$\approx$\,0.0146), while grains with $a_{\rm c}$\,$\approx$\,450\,$\mu\rm m$ ($\beta$\,$\approx$\,0.0047) dominate the emissions at $\lambda_{\rm obs}$ = 3000\,$\mu\rm m$. Fig. 1 shows that grains with radii $\sim$ 450\,$\mu\rm m$ are overabundant by a factor of 1.5 compared to grains with radii $\sim$ 145\,$\mu\rm m$ in the case of the very low material strength, eventually causing the observed lower spectral index in the case of the very low material strength. \newline\newline
\begin{figure*}
\centering
\includegraphics[width=18.5cm]{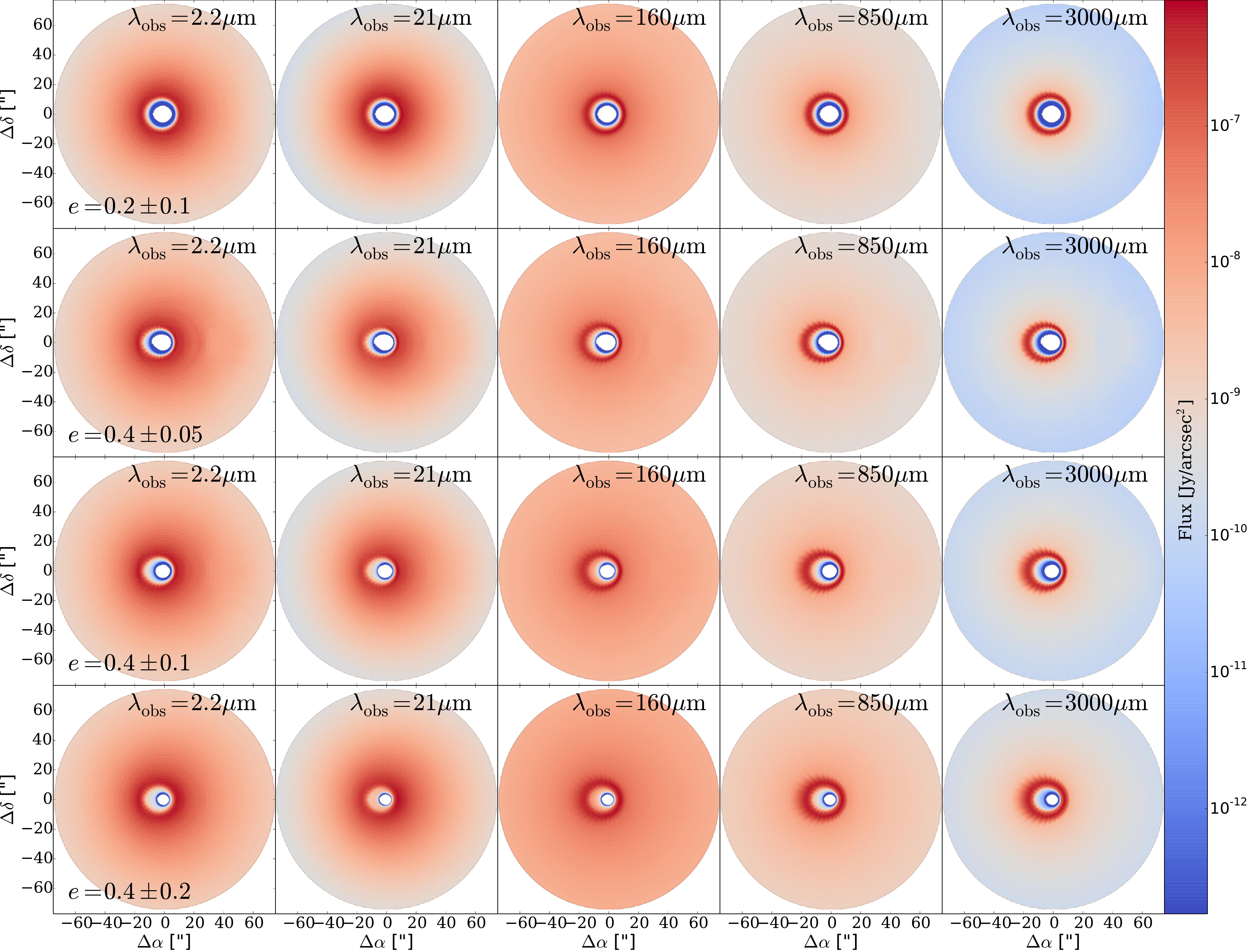}
\caption{The simulated surface brightness from near-IR to sub-mm wavelengths ($\lambda_{\rm obs}$ = 2.2\,$\mu\rm m$, 21\,$\mu\rm m$, 160~$\mu\rm m$, 850\,$\mu\rm m$, and  3000\,$\mu\rm m$), for different belt eccentricities ($e_{\rm b}$~=~0.2 and 0.4) and dynamical excitations ($\Delta$$e_{\rm b}$~=~0.05, 0.1, and 0.2) for the reference material strength. Both scattered and thermal re-emission are considered.}
\label{FigVibStab}
\end{figure*}
\begin{figure*}
\centering
\includegraphics[width=18cm]{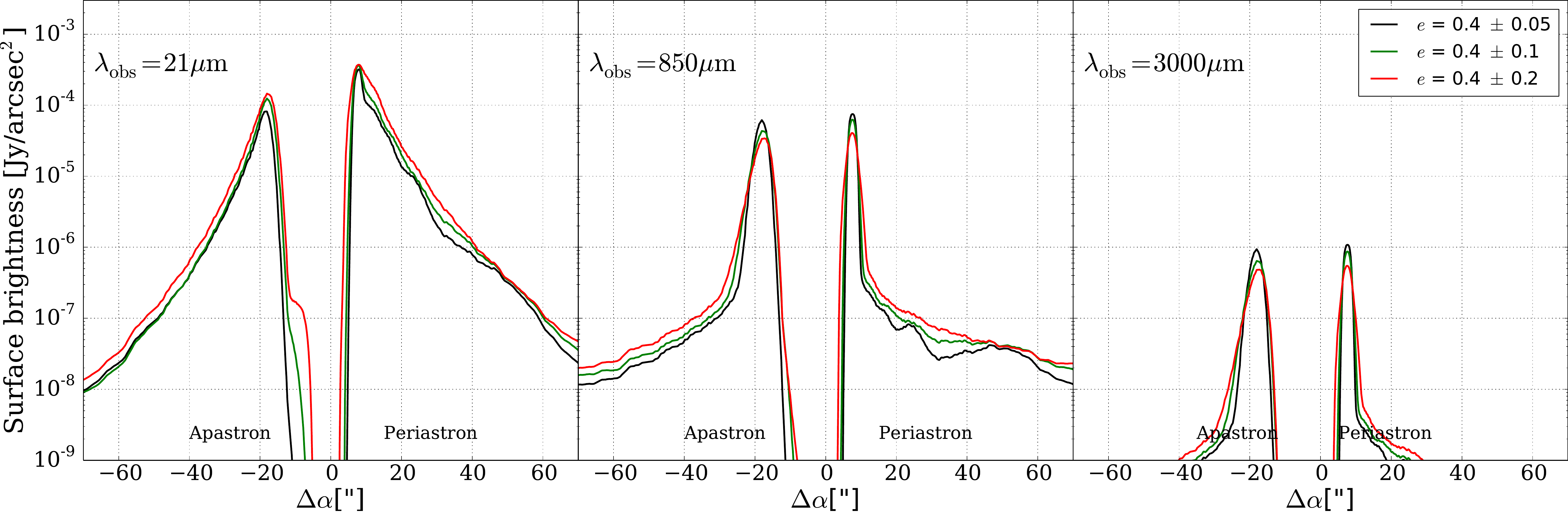}
\caption{The radial surface brightness profile as a function of dynamical excitations ($\Delta$$e_{\rm b}$) at wavelengths $\lambda_{\rm obs}$~=~21~$\mu\rm m$, 850\,$\mu\rm m$, and 3000\,$\mu\rm m$.}
\label{FigVibStab}
\end{figure*}
\subsubsection{Spatially resolved disk images}
\noindent In this section, we discuss impacts of collisional parameters, for instance, eccentricities $e_{\rm b}$, dynamical excitation $\Delta$$e_{\rm b}$, and the material strength $Q_{\rm s}$ on spatially resolved images and their radial surface brightness profiles from near-IR to sub-mm wavelengths. \newline\newline
{\noindent \bf 1. Dependence on the belt eccentricity and dynamical excitation} {\hspace{2mm}} The panels in Fig. 8 show simulated observations of spatially resolved disks at different belt eccentricities ($e_{\rm b}$~=~0.2 and 0.4), different levels of dynamical excitation ($\Delta$$e_{\rm b}$ = 0.05, 0.1, and 0.2) for the reference material strength $Q_{\rm s}$ at wavelengths $\lambda_{\rm obs}$ = 2.2\,$\mu\rm m$, 21\,$\mu\rm m$, 160\,$\mu\rm m$, 850\,$\mu\rm m$, and 3000\,$\mu\rm m$. Both scattered and thermal re-emission radiation are considered. Grains which contribute to the near-IR flux are the smallest within the entire distribution and thus most sensitive to the strong radiation pressure from the central star. Consequently, debris disks are featureless in this wavelength range. In contrast, the larger grains which are dominating the appearance (emission) at sub-mm wavelengths are only weakly affected by radiation pressure allowing the parent ring to be traced in this wavelength range.\newline
\indent In Fig. 9, radial surface brightness profiles are shown for different levels of dynamical excitation ($\Delta$$e_{\rm b}$ = 0.05, 0.1, and 0.2) of belt eccentricities $e_{\rm b}$ = 0.4 at wavelengths $\lambda_{\rm obs}$~=~21\,$\mu\rm m$, 850\,$\mu\rm m$, and 3000\,$\mu\rm m$, assuming the reference material strength. As outlined in Section 4.3.1, the abundance of the smaller particles on the periastron side is reduced. Therefore, the surface brightness is dominated by the large grains on the periastron side, whereas on the apastron side the smaller grains are the dominant population. Thus, if there are more destructive collisions, for instance, due to higher eccentricities or higher dynamical excitations, the production rate of small particles is increased on the apastron side (see Fig. 2). This situation is depicted in the simulated surface brightness profile at $\lambda_{\rm obs}$ = 21\,$\mu\rm m$. In the case of a higher dynamical excitation ($\Delta$$e_{\rm b}$ = 0.2) a brighter apastron side is observed than at lower dynamical excitation ($\Delta$$e_{\rm b}$ = 0.05). \newline
\indent The abundance of small dust grains is also subject to further forces and effects, including Poynting-Robertson effect and stellar drag forces (e. g., \citealp{Wyatt1999}; \citealp{Loehne2017}). For example, disks with a higher dynamical excitation show a higher production rate. Furthermore, smaller particles are most efficiently affected by Poynting-Robertson drag. Consequently,  the inner regions of dynamically excited debris disks appear more empty if observed at larger wavelengths as compared to observations at shorter wavelengths (tracing these smaller particles). In addition, Fig. 9 also shows that the higher dynamical excitation of the planetesimal belt come along with the increased overall brightness of the halo. \newline\newline
{\noindent \bf 2. Dependence on the material strength}{\hspace{2mm}} The panels in Fig. 10 show simulated observations of spatially resolved disks as a function of the material strength $Q_{\rm s}$ at wavelengths $\lambda_{\rm obs}$~=~2.2\,$\mu\rm m$, 21~$\mu\rm m$, 160\,$\mu\rm m$, 850\,$\mu\rm m$, and 3000\,$\mu\rm m$ ($e_{\rm b} = 0.4$). Both scattered and thermal re-emission radiation are considered. The different appearances are due to the different spatial and size distribution of the dust resulting from simulated collisional evolutions with different material strengths (Fig. 1). \newline
\begin{figure*}
\centering
\includegraphics[width=18.5cm]{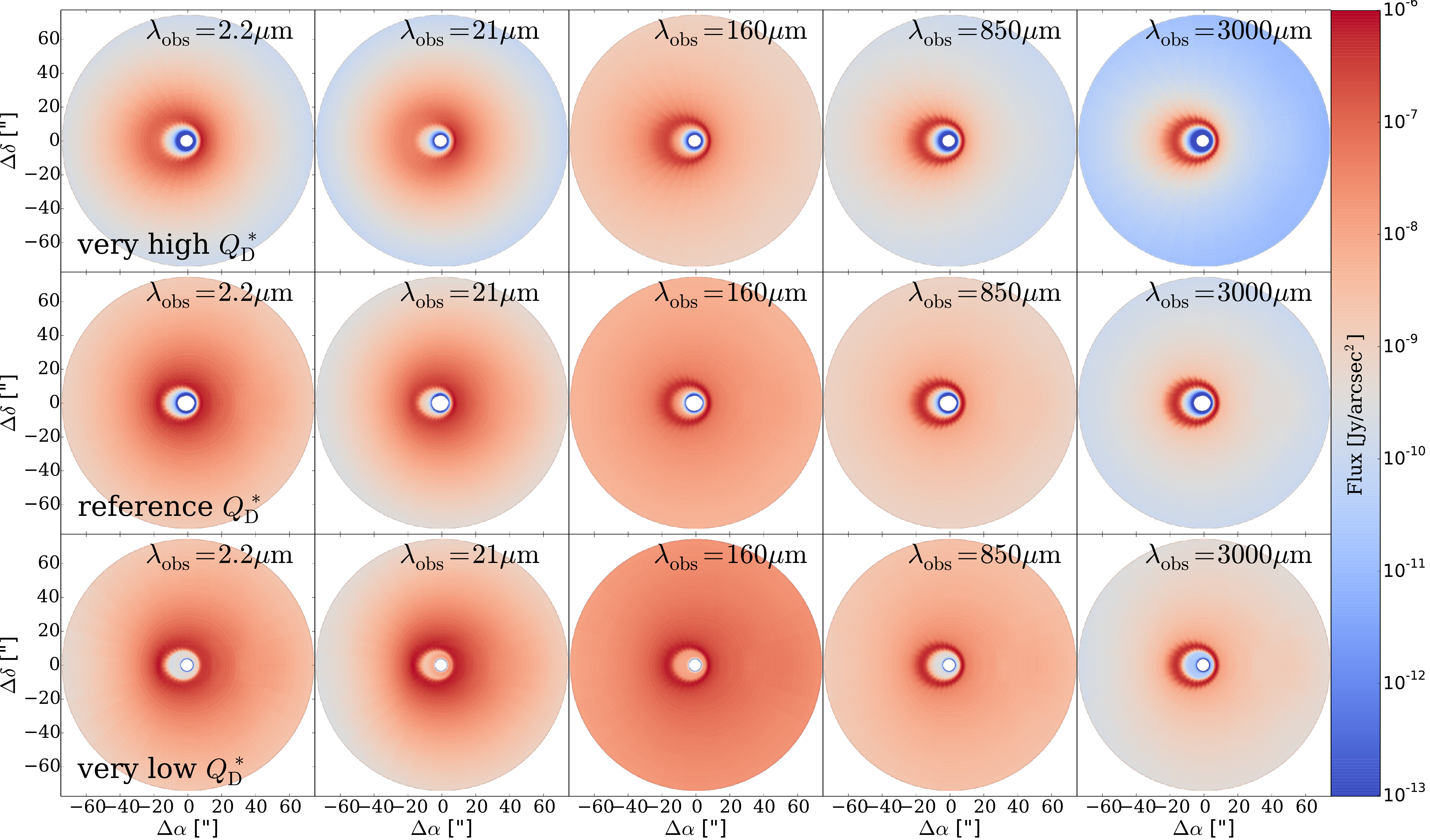}
\caption{The simulated surface brightness from near-IR to sub-mm wavelengths at $\lambda_{\rm obs}$ = 2.2\,$\mu\rm m$, 21\,$\mu\rm m$, 160\,$\mu\rm m$, 850\,$\mu\rm m$, and  3000\,$\mu\rm m$ (different material strengths $Q_{\rm s}$, fixed belt eccentricities $e_{\rm b}$ = 0.4). Both scattered and thermal re-emission are considered.}
\label{FigVibStab}
\end{figure*}
\begin{figure*}
\centering
\includegraphics[width=18cm]{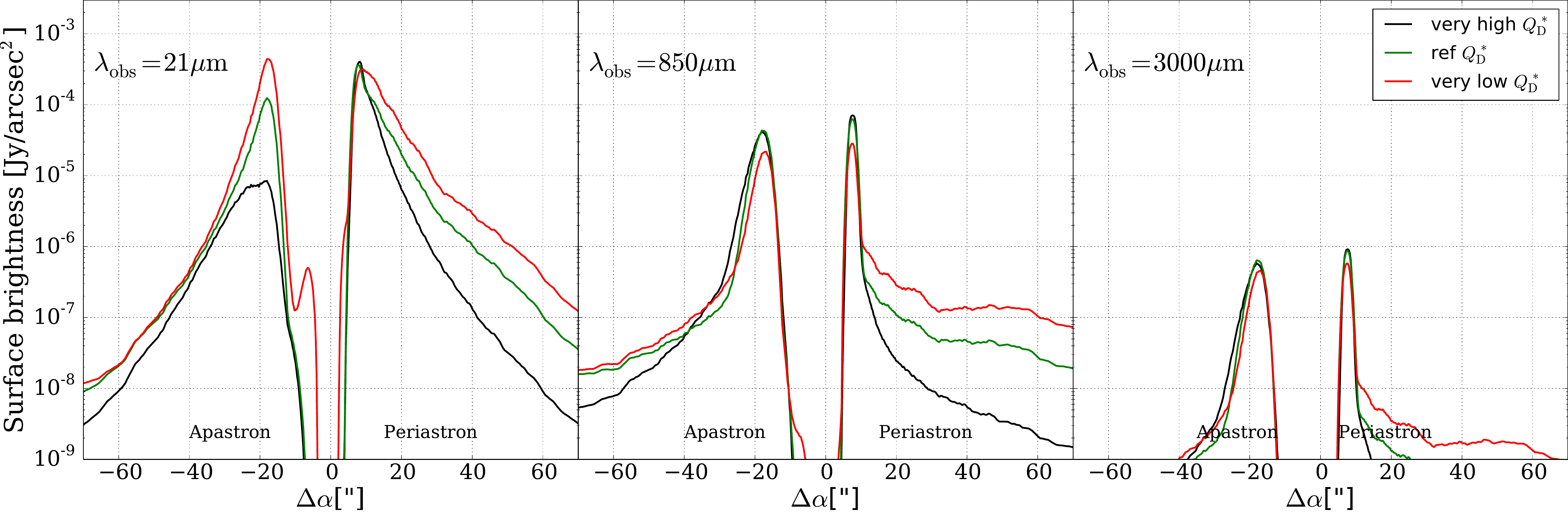}
\caption{The radial surface brightness profile as a function of the material strength; belt eccentricity $e_{\rm b}$ = 0.4; $\lambda_{\rm obs}$~=~21\,$\mu\rm m$, 850\,$\mu\rm m$, and 3000\,$\mu\rm m$; $\Delta \delta = 0$.}
\label{FigVibStab}
\end{figure*}
\indent As the periastron side of disks is closer to the star, one could expect that it is always brighter than the apastron side due to the higher dust temperatures and a higher rate of scattered radiation. However, as the particles travel more slowly and hence spend more time at the apocenter, an eccentric disk should be denser in this region (see Eq. 1). Furthermore, a very low material strength amplifies the increased rate of dust production. Thus, destructive collisions may reverse the situation of "pericenter glow" to an "apocenter glow". This can indeed be seen in Figs. 11, 12, and 13, where the brightness of the apastron side increases toward low material strengths. At sub-mm wavelengths ($\lambda_{\rm obs}$~$\ge$ 0.5~mm), far from the wavelength of the maximum emission, the impact of the dust temperature on the brightness distribution of the disk is reduced. Consequently, the effect of the pericenter glow is reduced. Instead, the impact of the dust number density on the surface brightness becomes more important. A complementary effect has been recently studied by \citet{Pan} and \citet{MacGregor}. They found that azimuthal temperature asymmetries, due to disk offset, could be compensated by azimuthal asymmetries in the dust density. \newline 
\indent In contrast, in the case of an increase of the material strength, the grain size distribution is critically changed by collisions (see Fig. 1). Thus, for very high material strengths, the surface brightness differences (asymmetries) are increased. Consequently, the pericenter glow phenomenon is even more pronounced for higher material strengths. \newline
\begin{figure*}
\centering
\includegraphics[width=18cm]{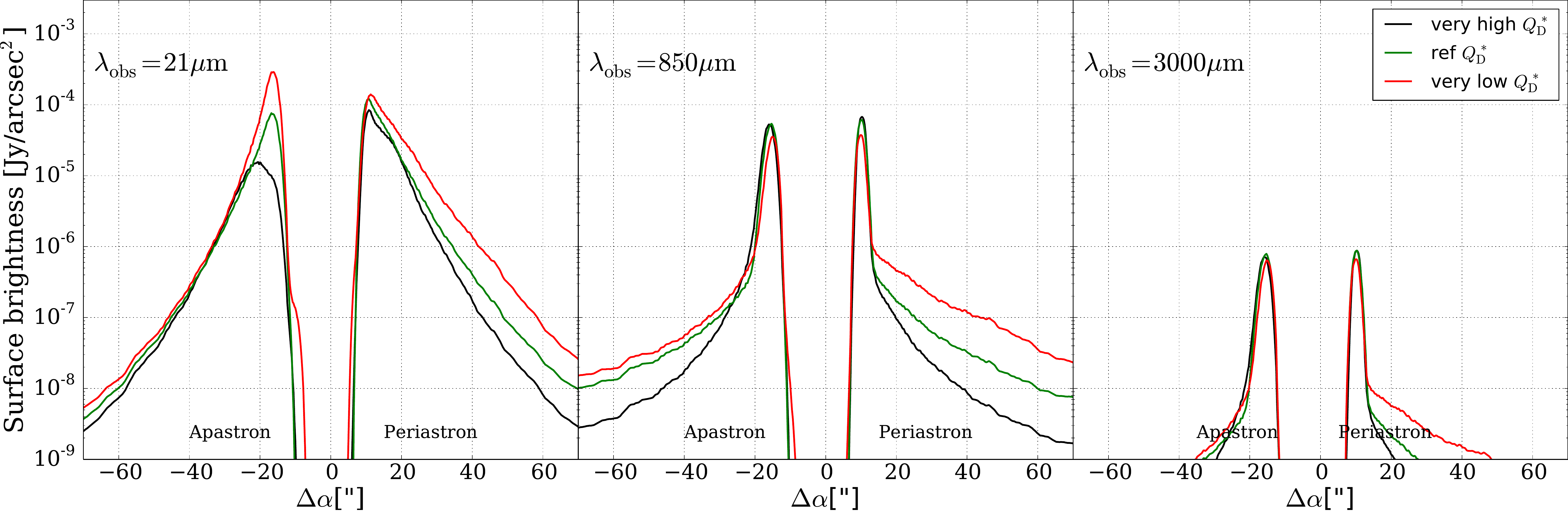}
\caption{The radial surface brightness profiles as a function of material strength; belt eccentricity $e_{\rm b}$ = 0.2; $\lambda_{\rm obs}$~=~21\,$\mu\rm m$, 850\,$\mu\rm m$, and 3000\,$\mu\rm m$; $\Delta \delta = 0$.}
\label{FigVibStab}
\end{figure*}
\begin{figure}
\centering
\includegraphics[width=9cm]{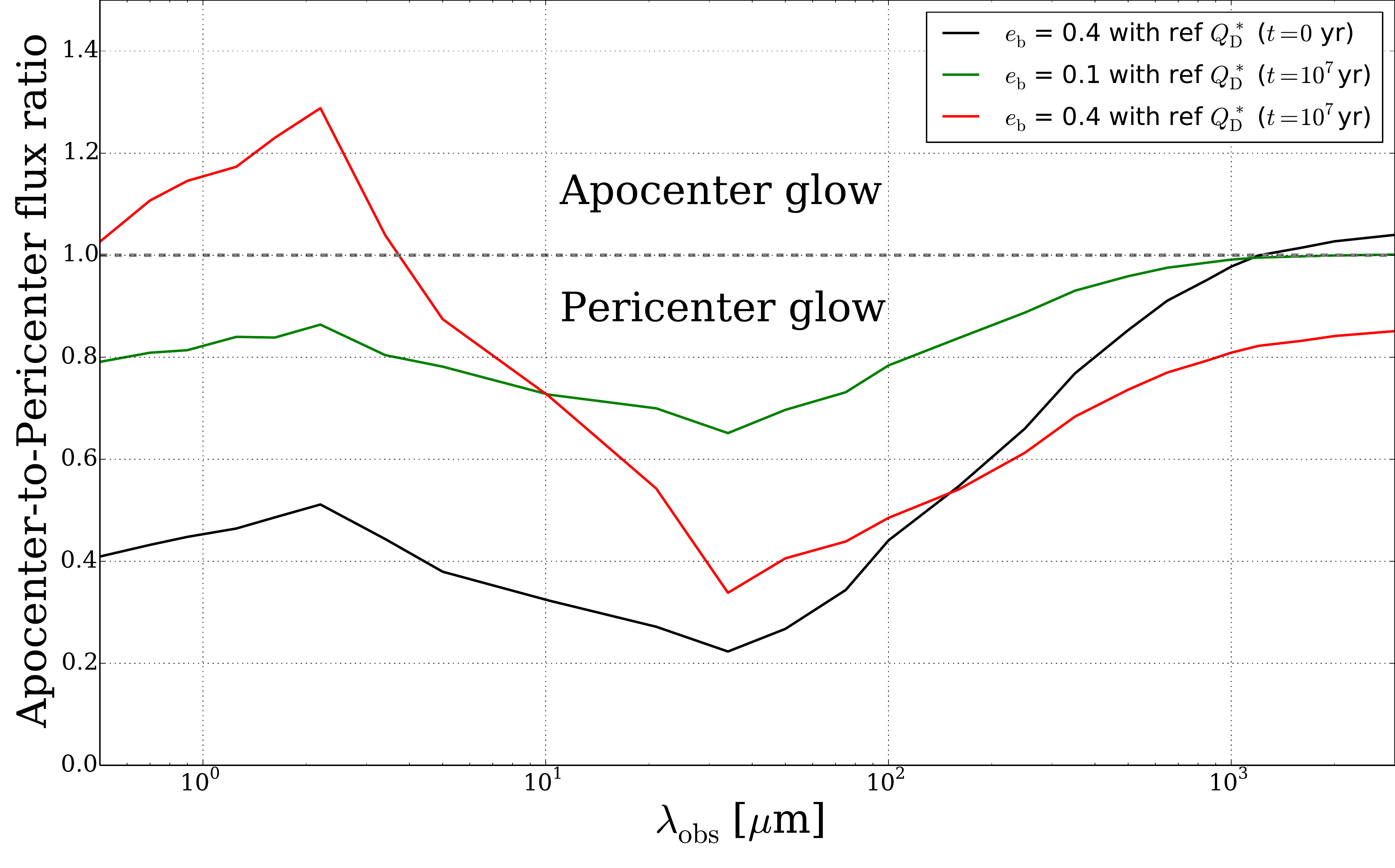}
\caption{The wavelength-dependence of the apocenter-to-pericenter flux ratio as a function of eccentricity. Black line: no collisions (t = 0 yr) with $e_{\rm b}$ = 0.4, ; green line: collisional evolution (t = 10$^{7}$ yr) $e_{\rm b}$ = 0.1; red line: collisional evolution (t = 10$^{7}$ yr) $e_{\rm b}$ = 0.4.}
\label{FigVibStab}
\end{figure}
\indent At shorter observing wavelengths $\lambda_{\rm obs}$, different material strengths $Q_{\rm s}$ seem only to result in a change of the brightness distribution at the apocenter (see Fig. 11), while the brightness distribution at the pericenter remains almost unchanged. However, the comparison at large distances that is the halo, different material strengths result in a significant change of the brightness distribution in the halo on the periastron side. At longer observing wavelengths $\lambda_{\rm obs}$, the impact of collisions becomes less important. Only the halo regions show differences in the brightness distribution. This behavior is in agreement with the findings outlined in Section. 4.3.2.: higher collisional rate results in an increased overall brightness of the halo. \newline
\indent At shorter observing wavelengths (see e.g., left panel of Fig. 11) we find again that smaller particles are abundant in the inner regions of the apastron sides due to the Poynting-Robertson drag force. In contrast, at longer observing wavelengths, inner regions appear void. In conclusion, we find that the higher collisional activity is responsible for symmetrizing their halo, changing the asymmetry of the peak of parent belts and even filling inner regions with a large fraction of smaller grains.\newline
\indent Fig. 12 shows the wavelength-dependent radial brightness profiles for different collisional evolutions for a belt eccentricity $e_{\rm b}$ = 0.2 at wavelengths $\lambda_{\rm obs}$ = 21\,$\mu\rm m$, 850\,$\mu\rm m$, and 3000\,$\mu\rm m$. We find that the effects observed for the belt eccentricity $e_{\rm b}$ = 0.4 and lower material strength, for instance, the apocenter glow, are still present and more pronounced in the disk with a belt eccentricity $e_{\rm b}$~=~0.2 and lower material strength. 
\subsection{Constraining collisional parameters from observational quantities}
\begin{figure*}
\centering
\includegraphics[width=18cm]{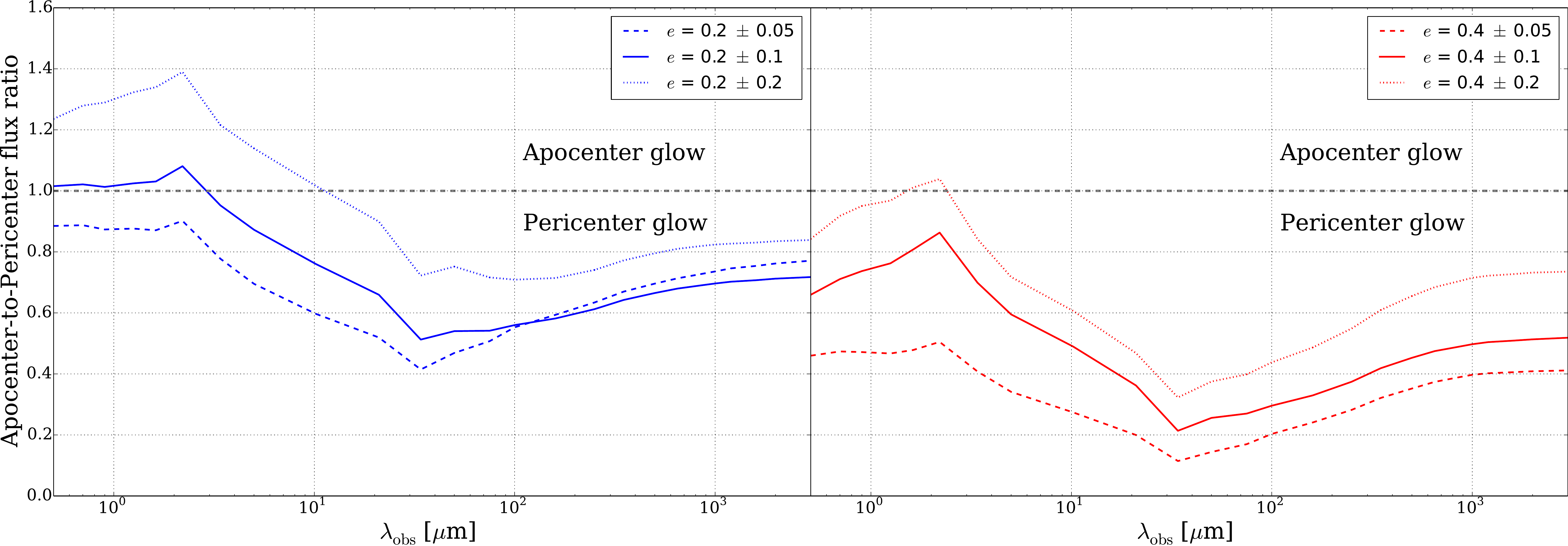}
\caption{The wavelength-dependence of the apocenter-to-pericenter flux ratio as a function of dynamical excitation ($\Delta e_{\rm b}$ = 0.05, 0.1, and 0.2) for eccentricities $e_{\rm b}$ = 0.2 (left) and $e_{\rm b}$ = 0.4(right)}
\label{FigVibStab}
\end{figure*}
\begin{figure}
\centering
\includegraphics[width=9cm]{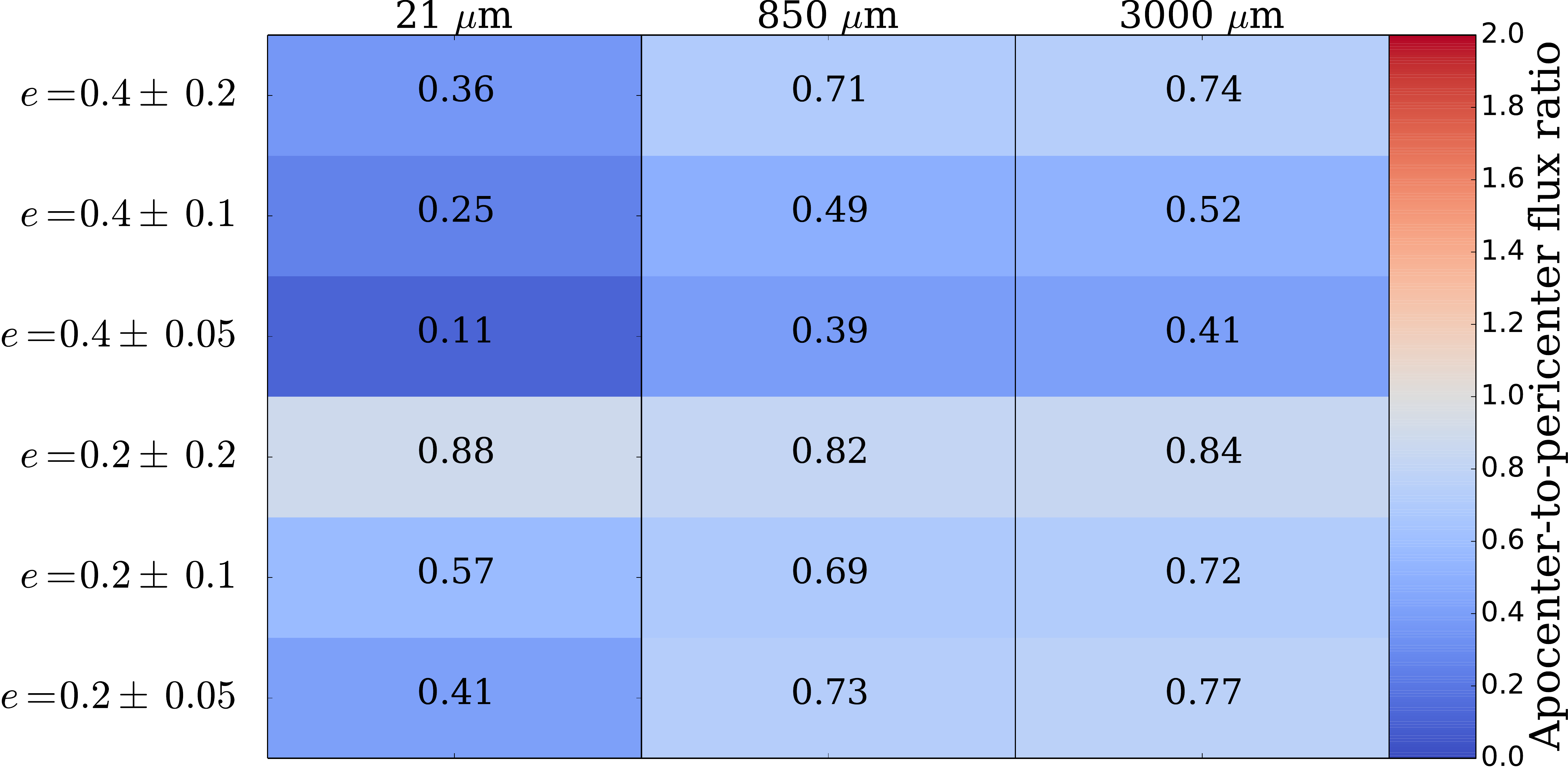}
\caption{The apocenter-to-pericenter flux ratio for different observing wavelengths $\lambda_{\rm obs}$ (21\,$\mu$m, 850 \,$\mu$m, and 3000~\,$\mu$m) as a function of dynamical excitation $\Delta e_{\rm b}$ (reference material strength). The number in each box indicates the apocenter-to-pericenter flux ratio.}
\label{FigVibStab}
\end{figure}
\begin{figure*}
\centering
\includegraphics[width=18cm]{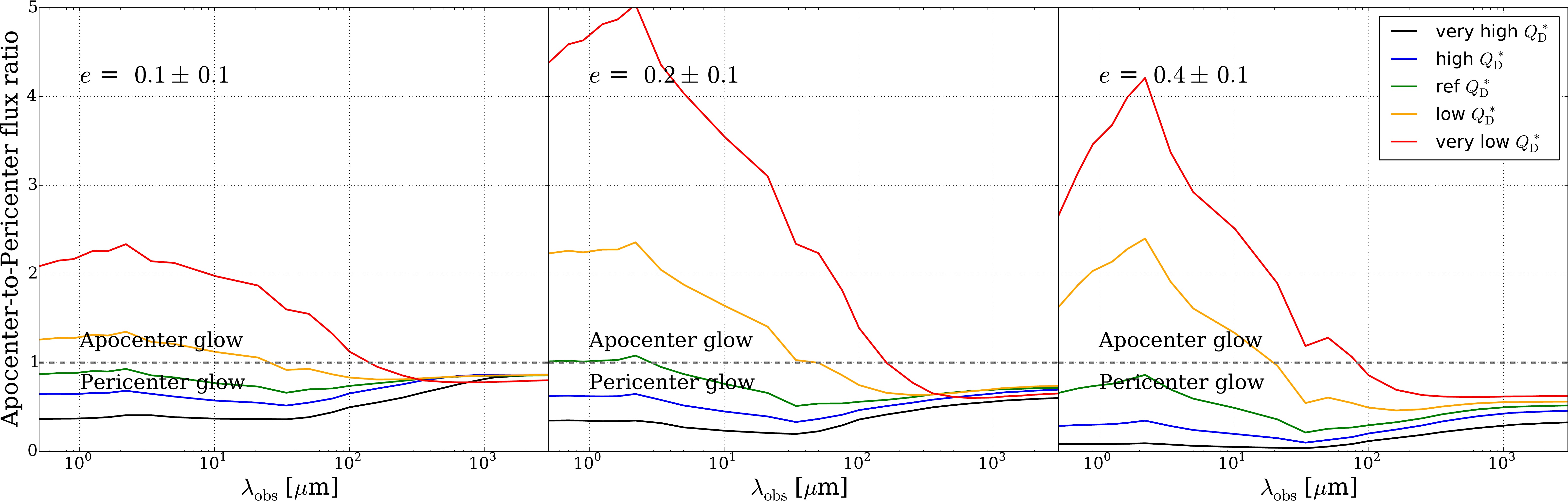}
\caption{The wavelength-dependence of the apocenter-to-pericenter flux ratio as a function of eccentricity $e_{\rm b}$ and material strength $Q_{\rm s}$.}
\label{FigVibStab}
\end{figure*}
\begin{figure*}
\centering
\includegraphics[width=18cm]{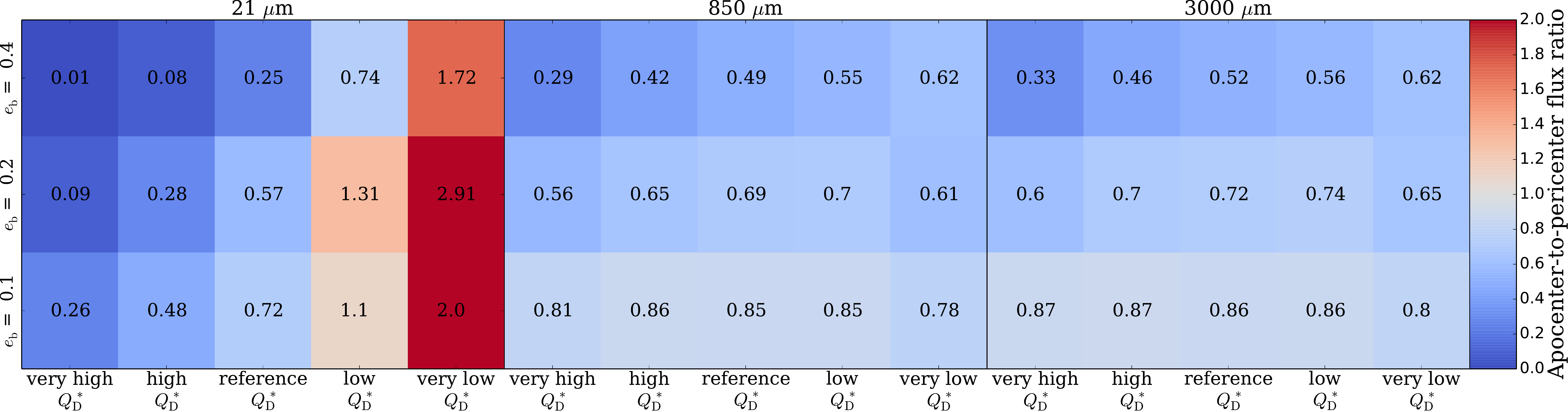}
\caption{The wavelength-dependence of the apocenter-to-pericenter flux ratio for selected values of the eccentricity $e_{\rm b}$ and material strength $Q_{\rm s}$. The number in each box indicates the apocenter-to-pericenter flux ratio. The case of the apocenter-to-pericenter flux ratio > 1 is discussed as the apocenter glow phenomenon in Sect. 4.4.}
\label{FigVibStab}      
\end{figure*}
\noindent In this section, we summarize the potential to constrain the collisional parameters from the observational quantities of debris disks. \newline\newline
\noindent \textbf{1. Belt eccentricity $e_{\rm b}$} {\hspace{2mm}} Fig. 13 depicts the apocenter-to-pericenter flux ratio as a function of wavelength for systems before and after the collisional evolution. We find that collisions have the potential to reduce the pericenter glow at shorter wavelengths. Furthermore, azimuthal asymmetries in the surface brightness at sub-mm wavelengths, which result from an increased dust density at the apastron side (\citealp{Pan}; \citealp{MacGregor}), are decreased due to collisions. \newline
\indent For two different belt eccentricities ($e_{\rm b}$ = 0.1 and $e_{\rm b}$~=~0.4) but the same level of dynamical excitation ($\Delta e_{\rm b}$ = 0.1), the situation is more complicated. The apocenter-to-pericenter flux ratio in the case of $e_{\rm b}$ = 0.1 is nearly constant for all wavelengths. In contrast, the apocenter-to-pericenter flux ratio for $e_{\rm b}$~=~0.4 is varying, which is due to the different size distributions at the apocenter and pericenter. For $e_{\rm b}$ = 0.1 the grain size distribution at periastron and apastron is much shallower and broader than for $e_{\rm b}$ = 0.4. This is not only due to the large number of smaller dust grains that survive at the apastron side of debris disks. It is also because more destructive collisions lead to an over-density of smaller particles and a depletion of larger particles. As a result, the pericenter glow phenomenon is reduced at shorter observing wavelengths and increased at longer observing wavelengths in debris disks with higher belt eccentricities (e.g., $e_{\rm b}$ = 0.4).\newline\newline
\noindent \textbf{2. Dynamical excitation $\Delta e_{\rm b}$} {\hspace{2mm}} Fig. 14 and 15 depict the wavelength-dependent apocenter-to-pericenter flux ratio for different levels of dynamical excitation ($\Delta e_{\rm b}$ = 0.05, 0.1, and 0.2) of different belt eccentricities ($e_{\rm b}$ = 0.2 and 0.4). For the case of $e_{\rm b}$ = 0.2, we find the apocenter glow phenomenon at shorter wavelengths in the case of the reference and higher dynamical excitation level ($\Delta e_{\rm b}$~=~0.1 and 0.2). The main reasons for this behavior are the proximity of the star to the apastron side and a higher rate of dust production due to higher relative velocities compared to the case of lower dynamical excitation level. For the case of $e_{\rm b}$ = 0.4, the minimum of the apocenter-to-pericenter flux ratio is smaller than in the case of the disk with $e_{\rm b}$ = 0.2 disks that is the more eccentric belt shows a stronger pericenter glow. The temperature dependence due to the proximity to the star is dominating this behavior over the higher production rate of smaller particles. \newline
\indent At short wavelengths, we find again a decrease of the pericenter glow phenomenon with increasing $\Delta e_{\rm b}$ (see also Fig. 9). At long wavelengths, the basic trend is similar but we find that the spread in observed flux ratios is much narrower. In the less dynamically excited case, for instance, $\Delta e_{\rm b}$~=~0.05 for $e_{\rm b}$~=~0.2, higher densities on the apastron sides can weaken the pericenter glow phenomenon. In conclusion, collisional parameters can be best constrained through observations of highly eccentric belts at short wavelengths\newline\newline
\noindent \textbf{3. Material strength $Q_{\rm s}$} {\hspace{2mm}} Fig. 16 and 17 show the wavelength-dependent apocenter-to-pericenter flux ratio for different material strengths $Q_{\rm s}$ and different belt eccentricities ($e_{\rm b}$ = 0.1, 0.2, and 0.4). First of all, we see that the material strength and eccentricity are degenerate to a large degree. For all eccentric parent belts ($e_{\rm b}$ = 0.1, 0.2, and 0.4), we find that a decrease of the pericenter glow phenomenon with the decreasing material strength. As described in Section. 4.3.2., the pericenter glow phenomenon is replaced by the apocenter glow phenomenon in the case of the lower material strengths. \newline
\indent Furthermore, the influence of the material strength is strongest at short wavelengths. The mid-IR images show the apocenter glow phenomenon as long as the material strength is below its reference value. Thus, it is clear that a smaller material strength results in the higher asymmetry in surface brightness and density. In contrast, the pericenter glow phenomenon is increased for the higher material strengths with decreasing eccentricity of the parent belts. At longer wavelengths (sub-mm/mm), the dominating grains are not significantly affected by the stellar radiation pressure anymore that is the apocenter-to-pericenter flux ratio becomes insensitive to details of the underlying collisional evolution. \newline



\section{Summary}
\noindent We studied impacts of collisional parameters, for instance, belt eccentricity $e_{\rm b}$, dynamical excitation $\Delta e_{\rm b}$, and the material strength $Q_{\rm s}$, on the observational appearance of eccentric debris disks. We have found surface brightness asymmetries that are caused by the combined effects of collisions and radiation pressurers. Our key results are as follows:\newline

\begin{enumerate}

	  \item An increasing belt eccentricity of the debris disk system leads to the survival of a large number of smaller dust grains on the apastron side. Thus the "pericenter glow" phenomenon is reduced with decreasing material strength at shorter observing wavelengths. However, the geometrical parameters describing the debris disks system that is the proximity of the star to the apastron side, must be considered as well.\newline
	
	  \item An increasing level of dynamical excitation in the eccentric parent belt of the debris disk system leads to a higher production rate of smaller particles and depletion of sub-mm grains. Consequently, the "pericenter glow" is reduced. \newline

      \item A decreasing material strength results in a higher production rate of smaller particles, which reduces the surface brightness differences between periastron and apastron. For very low material strengths, the pericenter glow phenomenon is reduced and eventually even replaced by the opposite effect, the apocenter glow in the near to mid-IR wavelength range. In contrast, an increasing material strength results in an increase of asymmetries in the surface brightness distribution.\newline

      \item The wavelength-dependent apocenter-to-pericenter flux ratio at short wavelengths is increased in the case of a higher efficiency of destructive collisions, caused by increasing dynamical excitation of the belt or decreasing material strengths. This effect is less pronounced at sub-mm wavelengths. Within the considered parameter space, the impact of the material strength $Q^{\rm *}_{\rm D}$ on the appearance of the disk is stronger than that of dynamical excitation of the belt.\newline
      
      \item The SED alone does not provide unique constraints on the collisional parameters considered. Thus, deriving unique constraints for the impact of considered collisional parameters based on the SED alone is hardly possible. \newline

\end{enumerate}     
\noindent Our study motivates the need to have a closer look at the impact of the specific type of dust on the presented results. For example, the specific roles of crystalline ice, amorphous ice, astrosilicates, PAH, and graphite deserve further investigations. \newline


\begin{acknowledgements}
    We would like to thank the anonymous referee for helpful and constructive suggestions and comments that greatly contributed to improving the final version of the paper. This work was supported by the Research Unit FOR 2285 "Debris Disks in Planetary Systems" of the Deutsche Forschungsgemeinschaft. MK, SW, and FK acknowledge the DFG for financial support under contracts WO 857/15-1. FK also acknowledges financial support through European Research Grant SNDUST 694520. TL and AVK acknowledge the DFG for financial support under contracts LO 1715/2-1 and KR 2164/13-1. 
\end{acknowledgements}


\nocite{*}
\bibliographystyle{aa}
\bibliography{bibliography}


\clearpage
\appendix

\section{\\DMS description}
\noindent The software tool \textbf{DMS} (\textbf{D}ebris disks around \textbf{M}ain-sequence \textbf{S}tars) has been developed for the simulation of the spectral energy distribution (SED) and wavelength- and inclination dependent scattered light images and polarization maps as well as thermal re-emission maps of an optically thin dust distribution. The key equations for calculation of scattered light intensity and polarization maps (as well as the corresponding contribution to the SED) are as follows:
\begin{align}
      &  W^{\rm sca}_{\rm\lambda}	         = L_{\rm\lambda, *} Q^{\rm sca}_{\rm\lambda} (a) \frac{\pi a^{\rm 2}}{4 \pi r^{\rm 2}} S_{\rm 11} (\theta) d\theta, \\
      & (W^{\rm sca}_{\rm\lambda})_{\rm pol} = L_{\rm\lambda, *} Q^{\rm sca}_{\rm\lambda} (a) \frac{\pi a^{\rm 2}}{4 \pi r^{\rm 2}} S_{\rm 12} (\theta) d\theta,
\end{align}
\noindent Here $L_{\rm\lambda, *}$, $W^{\rm sca}_{\rm\lambda}$ and $(W ^{\rm sca}_{\rm\lambda} )_{\rm pol}$ are the monochromatic luminosity, as well as the power of scattered and linearly polarized linearly scattered radiation of a spherical dust grain, respectively. The quantities $S_{\rm 11}$ and $S_{\rm 12}$ are elements of the M\"uller matrix, describing the scattering of radiation by the particle. The quantity $\theta$ denotes the scattering angle that is the angle between the directions of incident and scattered radiation. The dust particles are assumed to be a   spherical (radius $a$). \newline
\indent Thermal re-emission maps and contribution to the SED are calculated based on following equations:
\begin{align}
      & W^{\rm abs}   _{\rm\lambda} = L_{\rm\lambda, *} Q^{\rm abs}_{\rm\lambda} (a) \frac{\pi a^{\rm 2}}{4 \pi r^{\rm 2}},  \\
      & W^{\rm re-emi}_{\rm\lambda} = 4 \pi a^{\rm 2} Q^{\rm abs}_{\rm\lambda} (a) B_{\rm\lambda} (T_{\rm g}).
\end{align}
\noindent Here $W^{\rm abs}_{\rm\lambda},  W^{\rm re-emi}_{\rm\lambda}$ are the monochromatic power absorbed and re-emitted by a spherical dust grain with the absorption efficiency $Q^{\rm abs}_{\rm\lambda} (a)$ and temperature $T_{\rm g}$ at a distance $r$ from the star, respectively.\newline 
\indent The optical properties of the dust grains are computed using the software tool \textbf{miex} (\citealp{Wolf}), which is included in DMS.

\end{document}